\documentclass[11pt]{article}
\usepackage{jheppub}
\usepackage[utf8]{inputenc}
\usepackage{amsmath}
\usepackage{amssymb}
\usepackage{commath}
\usepackage[mathscr]{euscript}
\usepackage{color}
\usepackage{physics}
\usepackage{braket}
\usepackage{tensor}
\usepackage{soul}
\usepackage{hyperref}
\usepackage{multirow}
\usepackage[dvipsnames]{xcolor}
\usepackage{bclogo}
\usepackage{comment}
\usepackage{dsfont}
\usepackage{cancel}
\usepackage[normalem]{ulem} 
\usepackage[T1]{fontenc}
\usepackage{pifont}
\usepackage{bm}
\makeatletter
\def\@fpheader{\relax}
\makeatother

\numberwithin{equation}{section}

\setstcolor{red}
\begin{document}

\title{\vspace*{0.2cm} Pseudo-spontaneous $U(1)$ Symmetry Breaking in Hydrodynamics and Holography}

\author[1]{Martin Ammon,}
\author[2,3]{Daniel Are\'an,}
\author[4]{Matteo Baggioli,}
\author[1]{Se\'an Gray,}
\author[2,3]{Sebastian Grieninger}

\affiliation[1]{Theoretisch-Physikalisches Institut, Friedrich-Schiller-Universit\"at Jena,\\
Max-Wien-Platz 1, D-07743 Jena, Germany}

\affiliation[2]{Instituto de F\'isica Te\'orica UAM/CSIC, Calle Nicol\'as Cabrera 13-15, 28049 Madrid, Spain}
\affiliation[3]{Departamento de F\'isica Te\'orica, Universidad Aut{\'o}noma de Madrid, Campus de Cantoblanco, 28049 Madrid, Spain}
\affiliation[4]{Wilczek Quantum Center, School of Physics and Astronomy, Shanghai Jiao Tong University, Shanghai 200240, China \& Shanghai Research Center for Quantum Sciences, Shanghai 201315}
\preprint{IFT-UAM/CSIC-21-130}

\emailAdd{martin.ammon@uni-jena.de}
\emailAdd{daniel.arean@uam.es}
\emailAdd{b.matteo@sjtu.edu.cn}
\emailAdd{sean.gray@uni-jena.de}
\emailAdd{sebastian.grieninger@gmail.com}

\vspace{1cm}

\abstract{We investigate the low-energy dynamics of systems with pseudo-spontaneously broken $U(1)$ symmetry and Goldstone phase relaxation. We construct a hydrodynamic framework which is able to capture these, in principle independent, effects. We consider two generalisations of the standard holographic superfluid model by adding an explicit breaking of the $U(1)$ symmetry by either sourcing the charged bulk scalar or by introducing an explicit mass term for the bulk gauge field. We find agreement between the hydrodynamic dispersion relations and the quasi-normal modes of both holographic models. We verify that phase relaxation arises only due to the breaking of the inherent Goldstone shift symmetry.
The interplay of a weak explicit breaking of the $U(1)$
and phase relaxation renders the DC electric conductivity finite but does not result in a Drude-like peak.
In this scenario we show the validity of a universal relation, found in the context of translational symmetry breaking, between the phase relaxation rate, the mass of the pseudo-Goldstone and the Goldstone diffusivity.}

\setcounter{tocdepth}{2} 
\maketitle



\section{Introduction}

Symmetries play a fundamental role in our current understanding of Nature. According to Noether's theorem, each continuous symmetry is accompanied by a conserved symmetry current. In quantum field theory, in particular, the presence of symmetries is essential for the construction of effective field theories, see \cite{Penco:2020kvy} for a recent review. For instance, the conservation equations of symmetry currents form the basis of hydrodynamics -- the theoretical framework which captures the late-time, long-wavelength dynamics of massless degrees of freedom.

Nevertheless, realistic physical systems rarely exhibit exact symmetries, in which case a symmetry is said to be broken. There exists an assortment of mechanisms by which symmetries may be broken; we will concern ourselves with aspects of spontaneous symmetry breaking and explicit symmetry breaking. A symmetry is spontaneously broken when a ground state no longer preserves the global symmetries of its theory \cite{Beekman_2019}, while explicit symmetry breaking occurs when symmetry-violating terms appear in the equations which define a theory. Spontaneous symmetry breaking leaves the conservation equations intact, while explicit breaking renders the conservation of the corresponding current void.  

Spontaneous symmetry breaking finds applications in various domains of physics ranging from high energy to condensed matter physics \cite{Burgess:1998ku}. It is intimately connected with continuous phase transitions \cite{HOHENBERG20151} and hence serves as an effective tool in classifying different phases of matter -- the so called Landau paradigm -- which is especially useful from the perspective of effective field theory \cite{LVEFT3}. For continuous symmetries, a distinct feature of spontaneous breaking is the appearance of new massless degrees of freedom known as Goldstone bosons \cite{PhysRev.117.648,Goldstone1961}, a standard example of which is phonons in solids \cite{Leutwyler:1996er}. Several physical phenomena, such as superfluidity and elasticity, arise as a consequence of spontaneous symmetry breaking \cite{Lubensky}; in their hydrodynamic descriptions, the conservation equations are supplemented by the so-called Josephson relation, which governs the dynamics of the Goldstone boson.  

There is a long history of research dedicated to spontaneous symmetry breaking yet the study of its effects is still ongoing, particularly in connection with non-relativistic systems \cite{PhysRevLett.108.251602,PhysRevLett.110.091601}, spacetime symmetries \cite{PhysRevLett.88.101602}, open and out of equilibrium systems \cite{PhysRevE.97.012130,PhysRevD.103.056020}, topological symmetries and novel higher form symmetries \cite{PhysRevLett.126.071601,Hofman_2019}.

Physical situations arise where spontaneous symmetry breaking occurs in conjunction with explicit symmetry breaking; if the parameter causing the explicit breaking is small compared to the amount of spontaneous breaking (which is measured by the order parameter), the total breaking is said to be \textit{pseudo-spontaneous} \cite{PhysRevLett.29.1698}. In this scenario, the Goldstone bosons acquire a finite mass which follows the universal Gell-Mann-Oakes-Renner relation -- the mass-squared of such pseudo-Goldstone bosons is proportional to the explicit breaking scale. The quintessential example of pseudo-Goldstone bosons is pions as they are described in chiral perturbation theory \cite{PhysRev.175.2195}.

In the regime of pseudo-spontaneous symmetry breaking, the massive pseudo-Goldstone bosons and the violation of the conservation equations present difficulties for the application of hydrodynamics. However, for sufficiently  small explicit breaking and small pseudo-Goldstone mass one may expect that a system may still lend itself to a generalised hydrodynamic description \cite{nla.cat-vn2690514}. Such a regime enters the realm recently re-labelled as \textit{quasi-hydrodynamics} \cite{Grozdanov:2018fic}.

Besides explicit symmetry breaking, the Goldstone modes may be disrupted by a mechanism which is known as \textit{phase relaxation} --  a relaxation which is usually induced by topological defects. Phase relaxation may appear in systems which exhibit only spontaneous symmetry breaking. In the formulations of Goldstone dynamics based on higher-form symmetries, phase relaxation appears as an explicit breaking in a conservation law of a higher-form topological current \cite{Grozdanov:2016tdf,Grozdanov:2018ewh,Delacretaz:2019brr,BAGGIOLI20201,baggioli2021deformations}; this occurs independently of any explicit breaking of other symmetries of the system. The effects of phase relaxation are captured by a modified Josephson relation \cite{PhysRevB.96.195128}. Physical realisations of phase relaxation appear as elastic defects, for example dislocations and disclinations, which are responsible for melting in two-dimensional solids \cite{PhysRevLett.41.121}, and vortices in superfluids which relax the supercurrent leading to a finite electric DC conductivity \cite{PhysRev.140.A1197,Halperin1979,PhysRevB.94.054502}. 

Despite the principal separation between phase relaxation and pseudo-spontaneous symmetry breaking, it has been shown in the context of holography that the interplay between spontaneous and explicit symmetry breaking induces an effective phase relaxation rate \cite{Amoretti:2018tzw}. The effects of this effective phase relaxation have been examined in several different homogeneous \cite{Baggioli:2019abx,Ammon:2019wci,Andrade:2018gqk,Donos:2019txg,Donos:2019hpp,Amoretti:2021fch} and non-homogeneous \cite{Andrade:2020hpu} holographic models for translational symmetry breaking. Moreover, the appearance of a universal expression for effective phase relaxation, which relates it to the mass and diffusivity of the pseudo-Goldstone, has been noted and studied in the context of holographic models with translational symmetry breaking \cite{Amoretti:2018tzw,Ammon:2019wci,Baggioli:2019abx,Andrade:2018gqk,Donos:2019txg,Donos:2019hpp,Amoretti:2021fch}, effective field theories \cite{Baggioli:2020nay,Baggioli:2020haa} and the hydrodynamic description of QCD in the chiral limit \cite{Grossi:2020ezz}.

In this paper we will study pseudo-spontaneous breaking of $U(1)$ symmetry and phase relaxation through the means of hydrodynamics and holography. Ignoring the dynamics of the energy-momentum tensor we in section \ref{sec0} review superfluid hydrodynamics \cite{Herzog:2011ec} and incorporate explicit $U(1)$ symmetry breaking via charge relaxation and a mass for the pseudo-Goldstone, as well as phase relaxation. We find that such modifications affect the hydrodynamic modes, the retarded Green's functions and the behaviour of the AC and DC conductivity. We also re-derive a Gell-Mann-Oaks-Renner relation \cite{PhysRev.175.2195} for the mass of the pseudo-Goldstone by allowing for momentum dependent thermodynamic susceptibilities. 
In sections \ref{sec1} and \ref{sec2} we move on to holography and propose two distinct models -- stemming from the standard holographic superfluid \cite{Hartnoll:2008vx} -- for which the $U(1)$ symmetry of the dual theory is pseudo-spontaneously broken.
We consider these models in the probe limit.
In section \ref{sec1} we introduce a small source for a charged scalar operator in the bulk \cite{Argurio:2015wgr,Donos:2021pkk}; the dual field theory displays a massive pseudo-Goldstone boson as well as an effective phase relaxation which vanishes with the explicit breaking. Furthermore, the phase transition becomes smeared. We match our hydrodynamic theory for pseudo-spontaneous $U(1)$ symmetry breaking to the lowest quasi-normal modes.
We also propose and test a $U(1)$-appropriate version of the universal phase relaxation relation mentioned above.
In section \ref{sec2} we consider a finite mass term for the bulk gauge field, explicitly breaking the $U(1)$ symmetry \cite{Jimenez-Alba:2015awa,Jimenez-Alba:2014iia}. For small explicit breaking this model displays the effects of charge relaxation and the phase transition remains sharp. We suggest and test an expression for the charge relaxation in terms of the mass of the gauge field. We again match our hydrodynamic theory to the lowest quasi-normal modes.

Our findings may be of general relevance for chiral symmetry breaking in QCD and its incarnation at small but finite quark masses \cite{Son:2002ci,Grossi:2021gqi,Grossi:2020ezz,Florio:2021jlx}. Other possible extensions of our results regard the phenomenology of vortices in superfluids \cite{PhysRevB.94.054502} and ordered phases, such as charge-density waves \cite{RevModPhys.60.1129} in the context of translational symmetry breaking.
\\ \\
\textbf{Note added:} During the completion of this manuscript we became aware of the forthcoming work of \cite{Delacretaz:2021qqu}, which discusses the universal nature of the effective phase relaxation. Our results and those of \cite{Delacretaz:2021qqu} are in agreement, where applicable. 


\section{Hydrodynamics in the Probe Limit}
\label{sec0}

Hydrodynamics is an effective theory which describes the long-wavelength, late-time dynamics of a system. We will consider relativistic hydrodynamics and its applications to quantum field theory, where a spontaneously broken global $U(1)$ symmetry gives rise to superfluidity.

A superfluid can be viewed as a two-component fluid consisting of a normal component and a superfluid component \cite{RevModPhys.46.705,doi:10.1063/1.3248499}. Superfluids support propagating modes even when the dynamics of the normal component is frozen, i.e. when the energy-momentum tensor and fluctuations of the temperature and normal fluid velocity are neglected -- we will refer to this regime as the \textit{probe limit}. For simplicity and for applications to holography, we will only consider probe-limit hydrodynamics in this section -- some further results beyond the probe limit may be found in appendix \ref{sec:hydroApp}. The foundation of our hydrodynamic analysis stems from \cite{Herzog:2011ec},\footnote{The complete superfluid hydrodynamic framework of \cite{Herzog:2011ec} was successfully matched to holographic models in \cite{Arean:2021tks}.} which is briefly reviewed below; we then proceed to examine the results of adding a small explicit breaking of $U(1)$ symmetry to the superfluid system. We will consider a superfluid in two spatial dimensions. Greek indices span the full spacetime components while Latin indices cover the spatial components.   


\subsection{Superfluid Review}
The spontaneous breaking of $U(1)$ symmetry gives rise to a Goldstone boson $\varphi$ which is identified with the phase of the order parameter of the superfluid. A local $U(1)$ transformation shifts the Goldstone as $\varphi \mapsto \varphi + \lambda_g$ and it is hence necessary to define the gauge-invariant quantity 
\begin{equation}
    \xi_\mu \equiv \partial_\mu \varphi - A_\mu,
\end{equation}
where the external gauge field $A_\mu$ couples to the conserved $U(1)$ current and transforms as $A_\mu \mapsto A_\mu + \partial_\mu \lambda_g$. We fix $A_\mu=0$ in the following.

In the probe limit the conservation equation is simply that of $U(1)$ charge and it reads
\begin{align}\label{conservationProbeSuper}
    \partial_\mu\braket{ J^\mu} &= 0\,,
\end{align}
where $\braket{J^\mu}$ is the conserved $U(1)$ current. The Goldstone dynamics is governed by the Josephson relation which at ideal level is given by
\begin{equation}\label{josephson1}
    u^\mu \xi_\mu = -\mu,
\end{equation}
where $u^\mu = (1,0,0)$ is the equilibrium normal fluid velocity normalised such that $u^\mu u_\mu = -1$, and $\mu$ is the chemical potential.\footnote{We here assume that the superfluid chemical potential is identical to the normal fluid chemical potential.} The Josephson relation tells us that $\xi_\mu$ is to be considered as zeroth order in derivatives.

Up to first order in derivatives the constitutive relations and the spatial derivative of the Josephson relation read\footnote{Notice that the Josephson relation must be treated on equal footing with the constitutive relations and not with the conservation equations. This asymmetry can be lifted by considering a higher-form language \cite{Delacretaz:2019brr} where the Josephson relation is derived from the conservation of an additional emergent higher-form symmetry conservation law.}  
\begin{subequations}\label{eq:constrelationSuper}
\begin{align}
    \braket{J^{\mu}} &= \rho_t u^\mu + \rho_s n^\mu - \sigma_0 T \Delta^{\mu\nu} \partial_\nu \del{\frac{\mu}{T}} + \mathcal{O}(\partial^2),\label{constJmu}\\
        u^\mu \partial_\mu\del{\tensor{\Delta}{_\nu^\rho} \xi_\rho} &= \tensor{\Delta}{_\nu^\rho}\left[ -\partial_\rho \mu + \zeta_3 \partial_\rho \partial_\mu \del{\rho_s n^\mu}\right] + \mathcal{O}(\partial^2) , \label{josephsonDeriv}
\end{align}
\end{subequations}
where $\Delta^{\mu\nu}=u^\mu u^\nu + \eta^{\mu\nu}$ is the standard projector with $\eta^{\mu\nu}$ being the Minkowski metric. $T$ denotes the temperature while $\rho_n$ and $\rho_s$ respectively denote the normal and superfluid charge densities. The total charge density is given by $\rho_t =\rho_n + \rho_s$. The superfluid velocity $n^\mu$ is defined relative to the velocity of the normal component and is given by
\begin{equation}\label{supervelocity}
    \mu \,n^\mu \equiv \Delta^{\mu\nu}\xi_\nu\,.
\end{equation}
The transport coefficients entering in \eqref{eq:constrelationSuper} are the conductivity $\sigma_0$ and superfluid bulk viscosity $\zeta_3$ which, as evident from the Josephson relation \eqref{josephsonDeriv}, controls the Goldstone diffusivity; they are given by the Kubo formulae
\begin{subequations}\label{kuboSuper}
\begin{align}
    \sigma_0 &= -\lim_{\omega\to0}\frac{1}{\omega}\mathrm{Im}\left[G^\mathrm{R}_{J^x J^x}(\omega,0)\right] , \\
    \zeta_3 &= \lim_{\omega\to0}\omega\, \mathrm{Im}\left[ G^\mathrm{R}_{\varphi\varphi}(\omega,0) \right],\label{Gdiff}
\end{align}
\end{subequations}
where $G^\mathrm{R}_{ab}(\omega,\mathbf{k})$ is the retarded Green's function, $\omega$ denotes the frequency and $\mathbf{k}$ is the spatial momentum vector which will, without loss of generality, be taken along the $x$-direction such that $\mathbf{k}=(k,0)$. The thermodynamic and hydrodynamic quantities are functions of the chemical potential $\mu$ and temperature $T$.

We compute the spectrum of longitudinal excitations from the conservation equation \eqref{conservationProbeSuper} together with the constitutive relation \eqref{constJmu} and the Josephson relation \eqref{josephsonDeriv};\footnote{In the probe limit we do not consider fluctuations of the temperature $T$ and velocity $u^\mu$.} we find a pair of sound modes with dispersion relations
\begin{equation}\label{hydrodispProbe}
    \omega(k) = \pm v_s \,k - \frac{i}{2}~\Gamma_s \, k^2 + \dots,
\end{equation}
where the ellipsis denotes terms at higher order in $k$. This mode is historically called fourth sound \cite{Herzog:2011ec} to contrast it with the more standard second sound, see appendix \ref{sec:hydroApp}. The speed and attenuation of fourth sound are given by
\begin{align}
    v_s^2 = \frac{\rho_s}{\mu (\partial \rho_t / \partial \mu)}, \qquad \Gamma_s = \frac{\sigma_0}{(\partial \rho_t / \partial \mu)} + \zeta_3\frac{ \rho_s}{\mu}.
\end{align}

At this point a remark is in order. The probe limit superfluid hydrodynamics discussed in this section is not equivalent to considering the complete superfluid hydrodynamics in presence of momentum dissipation. In particular, our framework does not present the typical energy diffusion mode since the dynamics of the stress tensor $T^{\mu\nu}$ is not taken into account.


\subsection{Broken Superfluid}

When a symmetry is explicitly broken the current associated to the symmetry is no longer conserved. If the explicit symmetry breaking is sufficiently small and occurs in a background where the symmetry is also spontaneously broken the breaking is said to be pseudo-spontaneous and the system will contain massive (and damped) pseudo-Goldstone bosons \cite{Burgess:1998ku}. In this section the superfluid hydrodynamics presented above will be considered in a novel setting; namely, by including the effects of explicit $U(1)$ symmetry breaking the hydrodynamic framework will be appropriate for pseudo-spontaneous $U(1)$ symmetry breaking. To this end, we express the equation for the $U(1)$ charge current as
\begin{equation}\label{chargeNonCons}
    \partial_\mu\braket{J^\mu} = \Gamma u_\mu \braket{J^\mu} + m\,\varphi,
\end{equation}
where $\Gamma$ is the charge relaxation rate and $m$ is related to the mass of the pseudo-Goldstone $\varphi$.\footnote{For compactness we continue to use $\varphi$ to denote the pseudo-Goldstone boson.} We consider equation \eqref{chargeNonCons} at the level of fluctuations around thermal equilibrium. In order for the fluctuations of the charge current $J^\mu$ to be long-lived we in general assume $\Gamma$ and $m$ to be sufficiently small. 

An additional phenomenon which may affect the Goldstone bosons is phase relaxation, which enters the Josephson relation as a damping term and may appear independently of the explicit symmetry breaking. The relaxed superfluid Josephson relation takes the form 
\begin{equation}\label{relaxedJoseph}
    (u^\mu\partial_\mu + \Omega)\tensor{\Delta}{_\nu^\rho}\xi_\rho = \tensor{\Delta}{_\nu^\rho}\left[ -\partial_\rho\mu + \zeta_3 \partial_\rho \partial_\mu \del{\rho_s n^\mu}\right] + \mathcal{O}(\partial^2) ,
\end{equation}
where $\Omega$ denotes the phase relaxation rate, the inverse of the Goldstone lifetime. By considering the integrated version of the relaxed Josephson relation above, it becomes evident that $\Omega$ breaks the shift symmetry for $\varphi$. In the pseudo-spontaneous regime the thermodynamic and hydrodynamic quantities are dependent on the chemical potential, temperature and the explicit breaking scale.

\subsubsection{Hydrodynamic Modes}
Using the modified expressions \eqref{chargeNonCons} and \eqref{relaxedJoseph} together with the constitutive relation \eqref{constJmu}, we are able to determine the hydrodynamic modes using standard methods. The analysis yields a pair of modes with dispersion relations
\begin{equation}\label{gappedDiff}
    \omega(k) = \alpha_{\pm} - i D_{\pm} k^2 + \dots, \qquad \alpha_{\pm},D_{\pm} \in \mathbb{C}\,,
\end{equation}
where 
\begin{equation}\label{gap}
   \alpha_\pm = -\frac{i}{2}(\Gamma + \Omega) \pm \sqrt{\omega_0^2-\frac{(\Gamma - \Omega)^2}{4} }, \qquad \omega_0^2\equiv \frac{m}{(\partial \rho_t / \partial \mu)},
\end{equation}
with $\omega_0$ being the pinning frequency,\footnote{Leaving aside the definition of the pinning frequency, the structure of the gap \eqref{gap} is the same as to the case of pseudo-phonons as presented in \cite{PhysRevB.96.195128}.} and
\begin{align}\label{fufu}
    D_\pm &= \frac12  \del{  \frac{ \sigma_0}{(\partial \rho_t / \partial \mu) } + \zeta_3  \frac{\rho_s}{\mu} } \pm \frac{i}{2} { \frac{ \zeta_3  \rho_s (\partial \rho_t / \partial \mu)(\Gamma-\Omega) + 2\rho_s - \sigma_0 \mu ( \Gamma-\Omega)}{\mu\sqrt{4m(\partial \rho_t / \partial \mu) - (\partial \rho_t / \partial \mu)^2 (\Gamma-\Omega)^2 } } }.
\end{align}
The modes for the purely spontaneous phase are recovered by setting $\Gamma = \Omega = m = 0$ in the current-conservation equation and Josephson relation prior to expanding the dispersion relations at small wave-vector. Taking the limit directly from the equations above would result in a nonphysical divergent result which is a manifestation of the fact that the $k \rightarrow 0$ limit does not commute with the limit of zero explicit breaking. 
Furthermore, the dispersion relation may also be resummed and expressed in the form
    \begin{equation}\label{resum}
        \begin{split}
            \omega(k) &= \pm \sqrt{\frac{m}{(\partial \rho_t / \partial \mu)} - \frac{(\Gamma-\Omega)^2}{4} + k^2 \left( \frac{\rho_s}{\mu(\partial \rho_t / \partial \mu)} +(\Gamma-\Omega)\left( \zeta_3  \frac{\rho_s}{\mu} - \frac{ \sigma_0}{(\partial \rho_t / \partial \mu) }  \right)\right)} \\ 
        &\quad - \frac{i}{2} \left(\Gamma + \Omega \right) - \frac{i}{2}  \del{  \frac{ \sigma_0}{(\partial \rho_t / \partial \mu) } + \zeta_3  \frac{\rho_s}{\mu} } k^2 ,
        \end{split}
    \end{equation}
    for which the limit $\Gamma = \Omega = m = 0$ is well-defined and the relationship to the purely spontaneous phase is more evident. Expanding the square root in \eqref{resum} as a power series in $k$ yields the quantities displayed above at order $k^2$.

We note that the $k^2$-coefficients \eqref{fufu} sum to the same result in both the spontaneous and pseudo-spontaneous regimes, i.e.
\begin{equation}\label{sumruleprobe}
    D_+ + D_- = \Gamma_s.
\end{equation}
 An analogous `sum relation' also holds when the dynamics of the energy-momentum tensor is considered, see equation \eqref{sumruleApp} in Appendix \ref{sec:hydroApp}.
 
The result in equation \eqref{gap} allows for three different scenarios. First, if the expression under the square root in equation \eqref{gap} is negative the resulting modes will have a purely imaginary frequency at zero momentum, i.e. two damped relaxing modes. Moreover if either $\Gamma$ or $\Omega$ is zero in this regime one of the modes becomes undamped. Second, if the square root is positive the modes will acquire not only a finite damping (imaginary part at zero momentum) but also a real energy gap. Finally, in the third case a vanishing root results in the collision between the two modes on the imaginary axes.


\subsubsection{Retarded Green's Functions at \texorpdfstring{$\Gamma=0$}{gamma} }

We shall now present the retarded Green's functions of the system. In the following we set the charge relaxation rate to zero, $\Gamma = 0$. This is the configuration realised by the holographic setup in Section \ref{sec1}. We employ the canonical approach to retarded Green's functions with the conventions of \cite{2012}. The susceptibility matrix is given by
\begin{equation}
    \chi =
    \begin{pmatrix}
    \chi_{\rho\rho} & 0 \\
    0 & \tilde{\chi}_{\xi\xi}
    \end{pmatrix},
\end{equation}
where $\chi_{\rho\rho} = \partial\rho_t/\partial\mu$ and $\tilde{\chi}_{\xi\xi}$ is the susceptibility related to the pseudo-Goldstone boson. As a consequence of the explicit symmetry breaking, we allow for $\tilde{\chi}_{\xi\xi}$ to be momentum-dependent with the general parameterisation $ \tilde{\chi}_{\xi\xi} =\chi_{\xi\xi} f(k)$, where $\chi_{\xi\xi}$ is a positive constant.

For the retarded  Green's functions we find
\begin{subequations}
\begin{align}\label{eq:GF-offdiagonal}
 G^\mathrm{R}_{\varphi J^t}(\omega,k) &= \frac{i \omega \mu \chi_{\rho\rho} }{\mathfrak{p}(\omega,k)}, \\
      G^\mathrm{R}_{J^t \varphi}(\omega,k) &=\frac{i \omega (m \mu + k^2 \rho_s) \chi_{\xi\xi} \chi_{\rho\rho} f(k)}{\mathfrak{p}(\omega,k) }  ,
\end{align}
\end{subequations}
where $\mathfrak{p}(\omega,k)$ denotes the polynomial
\begin{equation}
    \mathfrak{p}(\omega,k) = m \mu  - \mu \chi_{\rho\rho} \omega (\omega + i \Omega) + k^2 \del{\rho_s - i \zeta_3 \rho_s \chi_{\rho\rho} \omega + \mu \sigma_0 (\Omega-i\omega)} + k^4 \zeta_3 \rho_s \sigma_0 \,.
\end{equation}
Using the symmetry $ G^\mathrm{R}_{\varphi J^t}(\omega,k) =  G^\mathrm{R}_{J^t \varphi}(\omega,k)$ we conclude that the pseudo-Goldstone susceptibility must be given by
\begin{equation}\label{eq:GMORrel}
    \tilde{\chi}_{\xi\xi} = \frac{k^2 \chi_{\xi\xi}}{k^2 + \mathfrak{m}^2},
\end{equation}
where 
\begin{equation}\label{eq:GMOR}
   \mathfrak{m}^2 =\chi_{\xi\xi}\, m    \, , \qquad \chi_{\xi\xi} = \frac{\mu}{\rho_s} \, .
\end{equation}
The result \eqref{eq:GMOR} is the Gell-Mann-Oakes-Renner relation with $\mathfrak{m}^2$ being the mass of the pseudo-Golstone boson, see for example \cite{Argurio:2015wgr,Amoretti:2016bxs}. Hence, for the static pseudo-Goldstone Green's function we obtain
\begin{equation}\label{eq:susceptcorrelation}
         G_{\varphi \varphi}(0,k) = \frac{ \chi_{\xi\xi}}{k^2 + \mathfrak{m}^2}.
\end{equation}

At $k=0$, we find the retarded Green's functions
\begin{subequations}\label{GFset1}
\begin{align}
     G^\mathrm{R}_{J^t J^t}(\omega,0) &=\frac{m \mu \chi_{\rho\rho}  }{m\mu - \mu \chi_{\rho\rho} \omega (\omega+i\Omega) } \, , \\
   G^\mathrm{R}_{\varphi J^t}(\omega,0) &=\frac{i \mu \chi_{\rho\rho} \omega }{{m\mu - \mu \chi_{\rho\rho} \omega (\omega+i\Omega) }}  \, ,
\end{align}
and also
\begin{align}
    \lim_{k\to0}\frac{1}{k}G^\mathrm{R}_{J^x J^t}(\omega,k) &= -\frac{\chi_{\rho\rho} \omega \del{\rho_s + \mu \sigma_0(\Omega - i\omega)} }{m\mu - \mu \chi_{\rho\rho}\omega(\omega + i\Omega) }, \\
   \lim_{k\to0}\frac{1}{k}G^\mathrm{R}_{J^x \varphi}(\omega,k) &= \frac{i \rho_s + \mu \sigma_0 \omega + \rho_s \chi_{\rho\rho}\omega \Omega/m }{m\mu - \mu \chi_{\rho\rho}\omega(\omega + i\Omega) }.\color{black}
\end{align}
\end{subequations}
Finally, the correlators \eqref{GFset1} allow us to compute the spatial current-current correlator; we find
\begin{subequations}\label{conductivity-ish}
\begin{align}
    G^\mathrm{R}_{J^x J^x}(\omega,0) &= \frac{\rho_s}{\mu} - i\,\sigma_0 \,\omega\,,\\
        G^\mathrm{R}_{J^x J^x}(0,k) &= \frac{m\rho_s}{m\mu+k^2 \rho_s};
\end{align}
\end{subequations}
which implies the following expression for the low-frequency AC conductivity,\footnote{We will continue to call $\sigma(\omega)$ the conductivity even though the $U(1)$ current is not conserved. At small explicit breaking this wording is, at least approximately, justified.}
\begin{equation}
    \sigma(\omega)\,=\lim_{k\to0}\frac{i}{\omega}\left[G^\mathrm{R}_{J^x J^x}(\omega,k) - G^\mathrm{R}_{J^x J^x}(0,k)\right] =\sigma_0. \label{cc}
\end{equation}
Although the denominators of all the retarded Green's functions in \eqref{GFset1} are the same, and contain the phase relaxation $\Omega$ and the parameter $m$, the charge non-conservation equation acts in such a way that both of these quantities drop out of the above result.

In order to gain clarity with regards to Kubo formulae it is beneficial to expand some of the above retarded Green's at low frequency,
\begin{subequations}
\begin{align}
     G^\mathrm{R}_{J^t J^t}(\omega,0) & =  -\chi_{\rho\rho} - \frac{i\omega}{m} \chi_{\rho\rho}^2 \Omega + \mathcal{O}(\omega^2), \\
   G^\mathrm{R}_{\varphi J^t}(\omega,0) &  =\frac{i\omega}{m} \chi_{\rho\rho}   - \frac{\omega^2}{m^2} \chi^2_{\rho\rho} \Omega+ \mathcal{O}(\omega^3);
\end{align}
\end{subequations}
hence we find the Kubo formula
\begin{equation}
    \Omega = \lim_{\omega\to0}\frac{m}{\omega \chi_{\rho\rho}^2}\mathrm{\Im}\left[  G^\mathrm{R}_{J^t J^t}(\omega,0) \right],\label{kubokubo}
\end{equation}
or similarly using $G^\mathrm{R}_{\varphi J^t}(\omega,0)$. The relation \eqref{conductivity-ish} provides the Kubo formula for $\sigma_0$ as 
\begin{equation}
    \sigma_0= -\lim_{\omega\to0} \frac{1}{\omega}\mathrm{Im} \left[ G^\mathrm{R}_{J^x J^x}(\omega,0)\right].
\end{equation}
Finally, using the transport coefficients above, $\zeta_3$ may be determined from
\begin{equation}\label{zetakubo}
   \lim_{\omega\to0}\frac{1}{\omega} \lim_{k\to0} \frac{\partial^2}{\partial k^2} \mathrm{Im} \left[ G^\mathrm{R}_{J^tJ^t}(\omega,k)\right] = -2\frac{\chi_{\rho\rho}^2}{m^2\mu} \del{\zeta_3 m \rho_s - \rho_s \Omega - \mu \sigma_0 \Omega^2} .
\end{equation}
In the next sections we will test this hydrodynamic framework with two different holographic models and draw important conclusions regarding the mechanism of phase relaxation induced by the pseudo-spontaneous symmetry breaking.


\section{Holographic Superfluid with a Sourced Charged Scalar}
\label{sec1}
A holographic model for s-wave superfluidity was constructed in \cite{Hartnoll:2008vx}, where the boundary operator $\mathcal{O}$ condenses below a critical temperature $T_\mathrm{c}$. In this section we generalise this model by sourcing the operator $\mathcal{O}$, resulting in an explicitly broken global $U(1)$ symmetry. The source may be tuned to be small such that the pseudo-spontaneous limit is achieved. We will investigate the properties of this model using numerical techniques. For simplicity we restrict ourselves to a $(2+1)$-dimensional field theory and we will work in the probe limit, where the backreaction of the gauge and scalar fields onto the metric is not taken into account. Therefore, our numerical results will be reliable only sufficiently close to the superfluid critical temperature $T_\mathrm{c}$ below which the scalar condensate forms.


\subsection{Holographic Setup}
We consider the bulk gravitational model of \cite{Hartnoll:2008vx} defined by the action
\begin{equation}
\label{eq:action}
S\,=\,
\int \dd^{4}x\, \sqrt{-g}\left[
-\frac{1}{4}F_{\mu\nu}F^{\mu\nu}-|D\psi|^2-M^2|\psi|^2\right]\,,
\end{equation}
where $\psi(z)$ is a complex
scalar with mass $M$, $F\equiv \dd A$ is the field strength tensor of the $U(1)$ gauge field, and the covariant derivative is defined as $D_\mu\equiv \partial_\mu -i\, q\, A_\mu $, where $q$ is the charge of the scalar operator $\mathcal{O}$ dual to the scalar field $\psi$. 

The bulk geometry is given by an $\mathrm{AdS}_4$ Schwarzschild black hole which in Eddington-Finkelstein coordinates reads
\begin{align}\label{eq:metric}
&\dd s^2=\frac{1}{z^2}\left[-u(z)\,\dd t^2-2\, \dd t \dd z+ \dd x^2+\dd y^2\right] \, , \qquad \textrm{with}\qquad  u(z)=1-z^3.
\end{align}
The AdS radius has been set to one and all bulk masses will be presented in units of that. Moreover, $\{t,x,y\}$ denote the coordinates of the $(2+1)$-dimensional Minkowski spacetime, while $z$ is the radial coordinate of $\mathrm{AdS}_4$ with conformal boundary located at $z=0$. The black hole horizon is defined by $u(z_h)=0$ with horizon radius $z_h=1$.

In order to describe the equilibrium state of the gravity theory we utilise the ansätze
\begin{equation}\label{ansatzScalar}
A=A_t(z)\,\dd t\,,\qquad \psi(z)=\psi_1(z)-i \,\psi_2(z)\,,
\end{equation}
for the gauge field and the scalar field. The temperature and the chemical potential of the dual field theory are given by
\begin{equation}
    T\,=\,-\frac{u'(1)}{4\pi}= \frac{3}{4\pi}\,,\qquad \mu\,=\, A_t(0)\,,
\end{equation}
where the value of the gauge field at the horizon is set to $0$ for convenience.

We set the mass of the scalar field to be $M^2=-2$ and we assume standard quantisation. Consequently, the boundary expansion for the scalar fields $\psi_i$ are given by
\begin{align}
    \psi_i(z)=\psi_i^{(l)}z+\psi_i^{(s)}z^2+\mathcal{O}(z^3)\,,\qquad
    (i=1,2)\,\label{eq:bdryexp}
\end{align}
where the superscripts $(l),(s)$ denote the leading and subleading terms, and the index $i$ distinguishes the fields entering in $\psi$.

The source $\lambda$, which we take to be real, of the dual operator is given by the leading term of the boundary expansion of $\psi$ -- we thus identify $\lambda = \psi_1^{(l)}$ as well as $\psi_2^{(l)}=0$. Next, in order to compute the expectation value we follow the procedure detailed in appendix A of \cite{Arean:2021tks}, arriving at
\begin{equation}
\langle \mathcal{O} \rangle =
2\psi_1^{(s)}+i\,2\psi_2^{(s)}
+i\,2 q \mu\, \psi_1^{(l)}=
2\,\psi_1^{(s)}\,,
\end{equation}
where in the second equality we have used the constraint
$\psi_2^{(s)}=-q \,\mu\,\psi_1^{(l)}$.

The case $\lambda=0$ is that of the original superfluid model of \cite{Hartnoll:2008vx} where the global $U(1)$ symmetry of the dual field theory is spontaneously broken. In the following we switch on an infinitesimal source $\lambda$ which will explicitly break the $U(1)$ symmetry; a larger value of $\lambda$ means more symmetry breaking.

The boundary Ward identity for the broken $U(1)$ symmetry reads
\begin{equation}\label{brokenWI}
    \partial_\mu \braket{J^\mu} = \frac{iq}{2}\left[\psi^{(l)} \braket{\mathcal{O}^*} - \left(\psi^{(l)}\right)^* \braket{\mathcal{O}} \right],
\end{equation}
where the asterisk indicates complex conjugation.\footnote{The unusual normalisation of the Ward identity is due to the factor of $2$ relating $\braket{\mathcal{O}}$ and $\psi^{(s)}.$} For real $\lambda$ also the expectation value of $\mathcal{O}$ is real at equilibrium such that right-hand side of the Ward identity \eqref{brokenWI} vanishes
and hence the current is conserved. Away from equilibrium, however, the right-hand side of the Ward identity \eqref{brokenWI} does not vanish; in particular, for the chosen real-valued source we obtain 
\begin{equation}
    \partial_\mu \langle J^\mu \rangle \,=\,\lambda\,q \, \mathrm{Im}\braket{\mathcal{O}} \, , \label{WI1}
\end{equation}
see also \cite{Amado:2009ts,Argurio:2015wgr,Donos:2021pkk}. Comparing the Ward identity \eqref{WI1} to the more general equation \eqref{chargeNonCons} one concludes that this holographic model has a vanishing current relaxation rate, $\Gamma=0$. Also, identifying $\varphi = \mathrm{Im}\langle{\cal O}\rangle/\langle{\cal O}\rangle_{\rm eq}$ the parameter $m$ relates to the source $\lambda$ as $m= q \lambda \braket{\mathcal{O}}_{\textrm{eq}}$, where $\braket{\mathcal{O}}_{\textrm{eq}}$ is the condensate in equilibrium. We will drop the subscript denoting the equilibrium in the following.

Fixing $q=1$, we use numerical techniques to determine the equilibrium condensate, the quasi-normal modes and the (retarded) Green's functions of the dual theory. For some details about the numerical methods see appendix \ref{app:num}.


\subsection{The Condensate and Goldstone Correlator}
\begin{figure}
    \centering
    \includegraphics[width=0.45\linewidth]{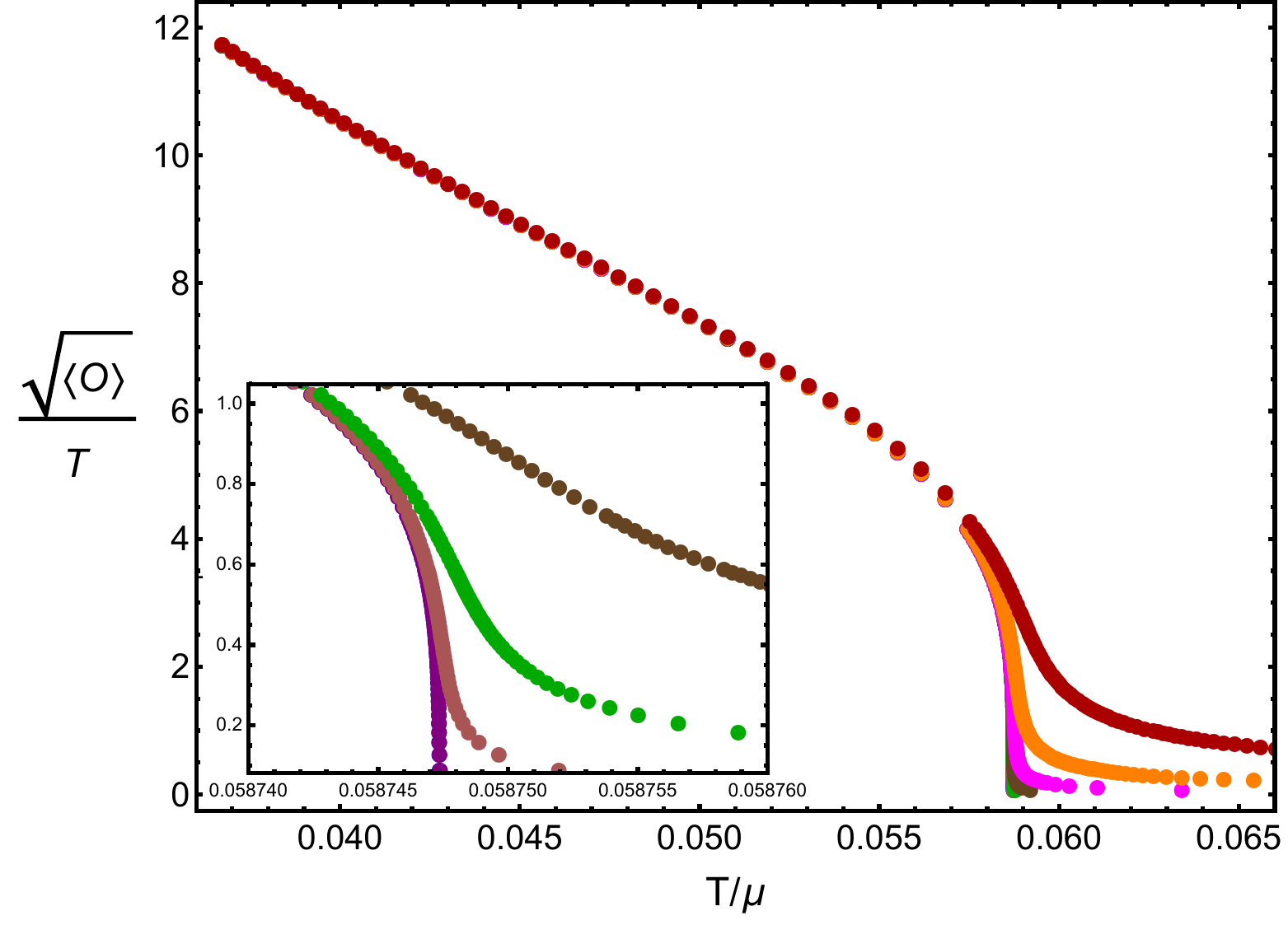}\quad  \includegraphics[width=0.45\linewidth]{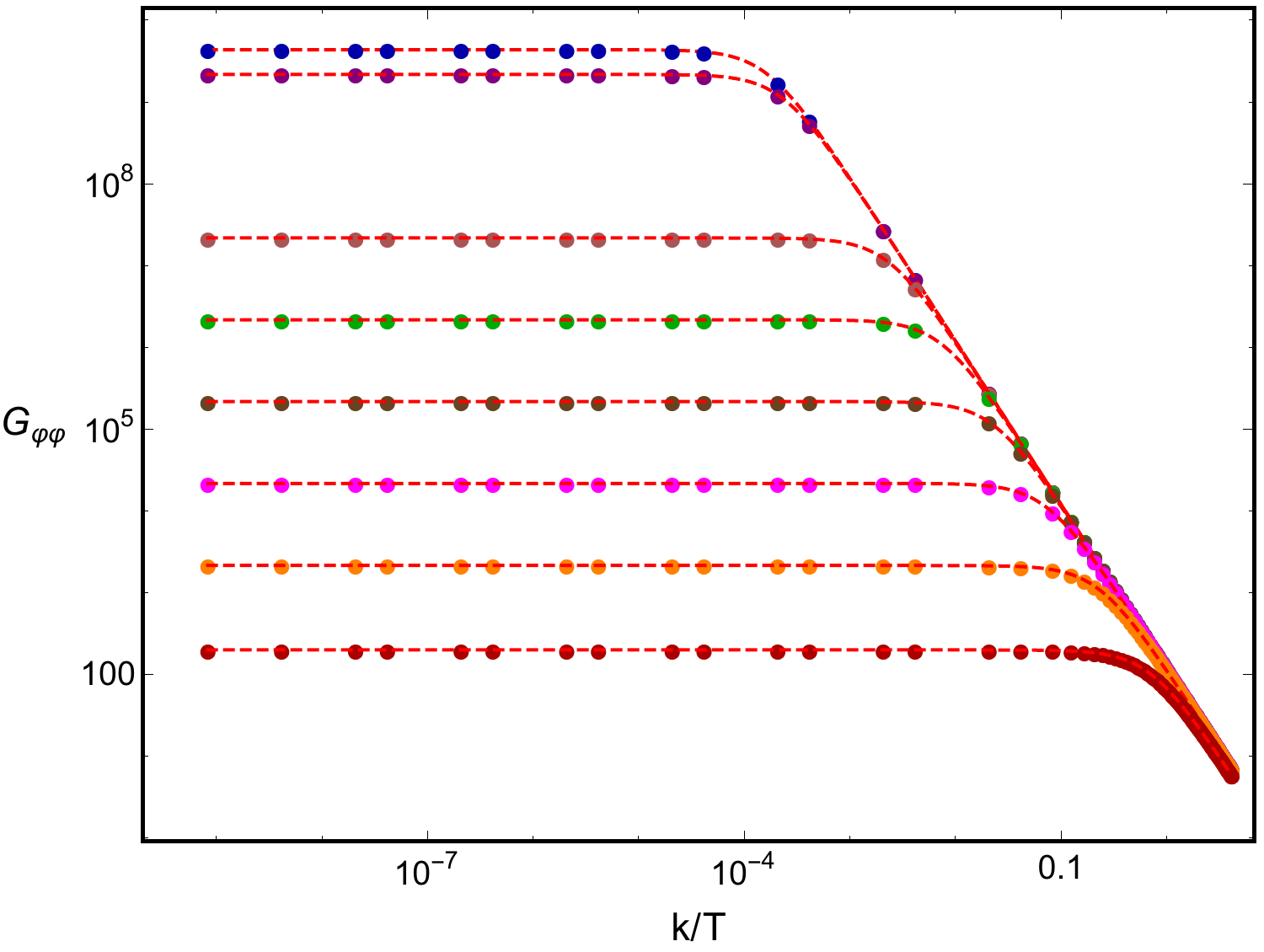}
    \caption{The colours indicate different values of the dimensionless source $\lambda/T$ -- blue is $5\times10^{-10}$, purple is $10^{-9}$, light brown is $10^{-7}$, green is $10^{-6}$, dark brown is $10^{-5}$, magenta is $10^{-4}$, orange is $10^{-3}$, and red is $10^{-2}$. \textbf{Left: }the dimensionless scalar condensate as a function of temperature close to the superfluid transition for different values of the dimensionless source $\lambda/T$. \textbf{Right: }the pseudo-Goldstone correlator at zero frequency and finite momentum at fixed $T/\mu=0.0575$ and for different values of the source $\lambda/T$. The dashed lines represent the expectations from hydrodynamics presented in the main text in equation \eqref{eq:susceptcorrelation}.}
    \label{fig1}
\end{figure}

We will first examine what effect a small source of explicit breaking has on the scalar condensate $\braket{\mathcal{O}}$ and the zero-frequency pseudo-Goldstone correlator $G_{\varphi\varphi}(0,k)$ given by \eqref{eq:susceptcorrelation}. The numerical results are shown in figure \ref{fig1}. The left panel shows that the value of the dimensionless condensate increases with the introduction of explicit breaking, while the phase transition goes from sharp to continuous.\footnote{This phenomenon might be related to the concept of imperfect bifurcations \cite{LIU2007573,gaeta1990bifurcation} and is similar to the effects of disorder on continuous phase transitions \cite{vojta2013phases,Arean:2015sqa,Ammon:2018wzb}. It would be interesting to make this point clearer in the future.} The tail of the continuous crossover increases with the source $\lambda$ such that the condensate becomes non-zero for every value of $T/\mu$.\footnote{We have verified that the probe-limit background can be reached by via a smooth limit from a backreacted setup.}
The behaviour of the zero-frequency pseudo-Goldstone correlator $G_{\varphi\varphi}(0,k)$ is displayed in the right panel of figure \ref{fig1} and shows perfect agreement with the hydrodynamic formula \eqref{eq:susceptcorrelation}.\footnote{$G_{\varphi\varphi}(0,k)$ follows from standard linear response analysis in holography once we identify $\varphi=\mathrm{Im}\langle\delta {\cal O}\rangle/\langle {\cal O}\rangle_{\rm eq}$ where $\delta {\cal O}$ stands for the fluctuations of the condensate.} This agreement holds only in the limit of sufficiently small explicit breaking, $\lambda/T \ll 1$; for larger explicit breaking the pseudo-Goldstone approximation is invalidated, rendering the hydrodynamics framework of section \ref{sec0} inapplicable.


\subsection{Zero-momentum Excitations}\label{sec:zeromomOmega}
\begin{figure}
    \centering
      \includegraphics[width=0.6\linewidth]{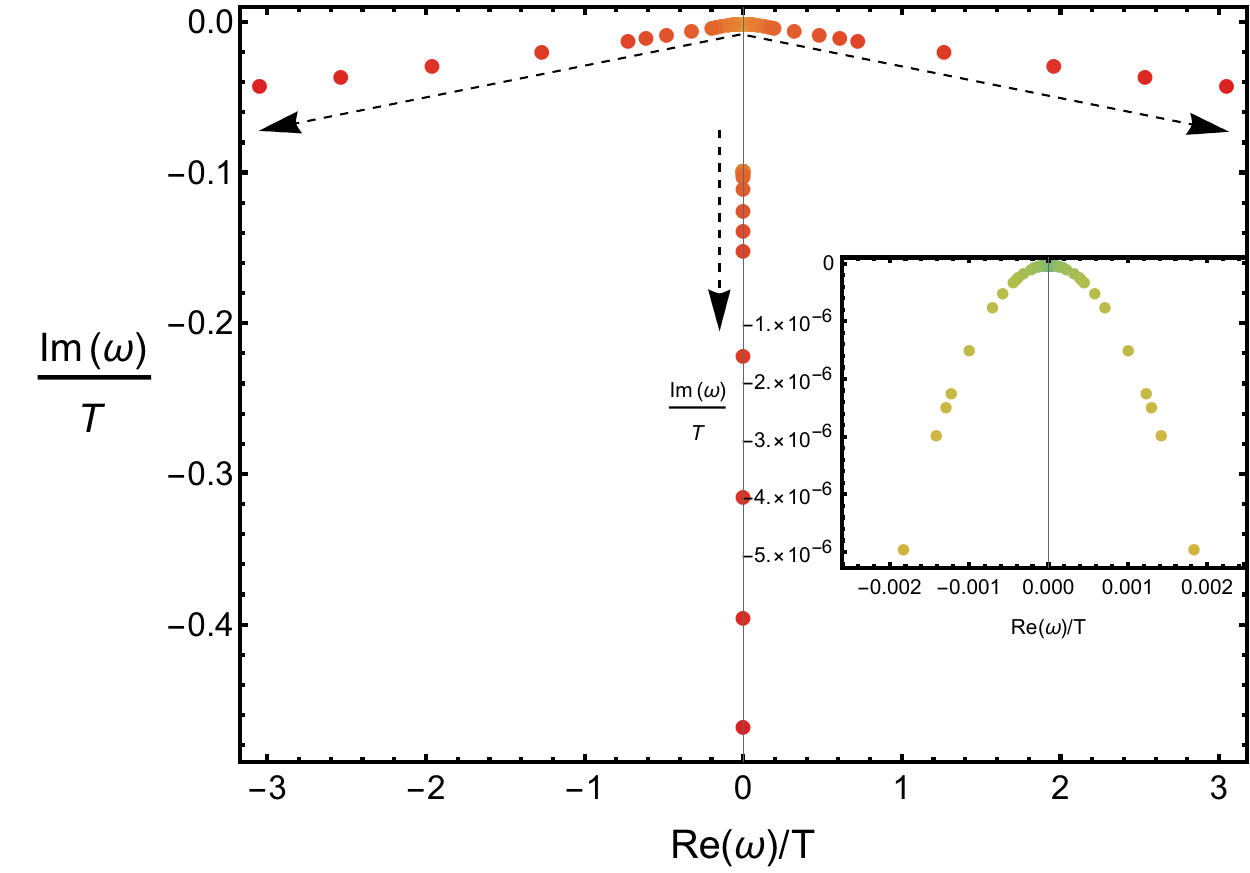}
    \caption{The lowest quasi-normal modes at zero wave-vector, in the complex plane, upon varying the source $\lambda/T$ in the range $[10^{-16},0.1]$ at fixed $T/\mu=0.0582$. The dashed arrows guide the eyes towards the limit of strong explicit breaking, $\lambda/T \gg 1$. The inset shows a zoom in the opposite limit of infinitesimal explicit breaking, $\lambda/T \ll 1$. We see a pair of sound modes which upon explicit breaking gain a complex gap, as well as a pseudo-diffusive mode which moves away from the origin as the breaking increases. }
    \label{fig2}
\end{figure}

The zero-momentum modes will be controlled by the expression in \eqref{gap}; in the absence of a charge relaxation rate, $\Gamma=0$, the dispersion relations read
\begin{equation}
    \omega = -i\,\frac{\Omega}{2}\pm \sqrt{ \omega_0^2- \frac{\Omega^2}{4} }\,,\qquad \omega_0^2\equiv \frac{m}{(\partial\rho_t/\partial \mu)}\,.\label{pred1}
\end{equation}
Numerical data showing the dynamics of the lowest quasi-normal modes of our model \eqref{eq:action} at zero momentum are displayed in figure \ref{fig2}. At zero breaking the spectrum contains a pair of sound modes sitting at the origin, and one non-hydrodynamic mode with finite imaginary gap.\footnote{Note that the gap of the non-hydrodynamic mode decreases as one approaches the phase transition.} When increasing the source of explicit breaking the non-hydrodynamic mode moves away from the origin along the imaginary axis; simultaneously, the modes which previously constituted sound modes acquire a finite complex gap, in accordance with equation \eqref{pred1} with a positive square root. For large values of the explicit breaking the modes with complex gap are further away from the origin than the mode with purely imaginary gap; in this regime a hydrodynamic description of the dynamics must fail \cite{Grozdanov:2018fic}.

\begin{figure}
    \centering
          \includegraphics[width=0.45\linewidth]{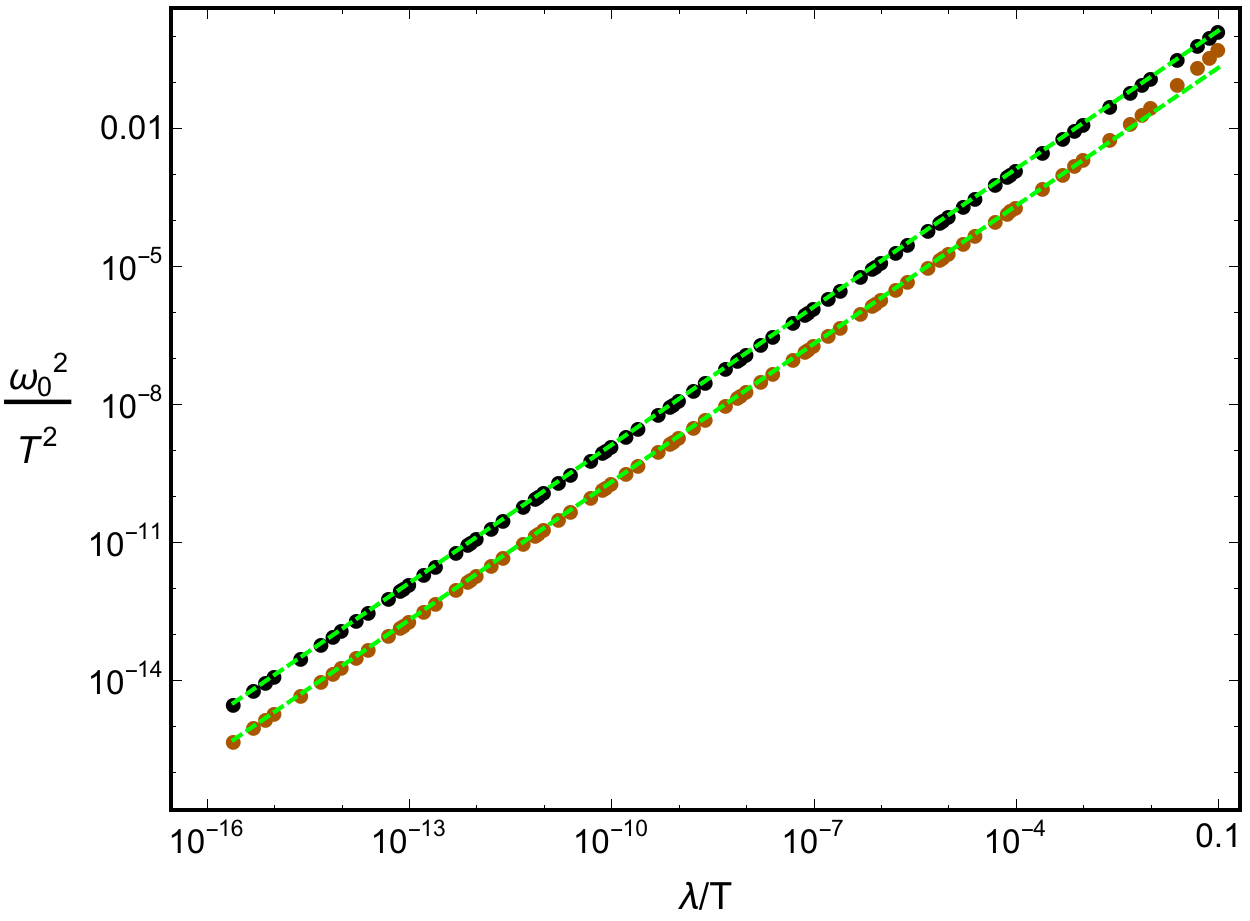}\quad
            \includegraphics[width=0.45\linewidth]{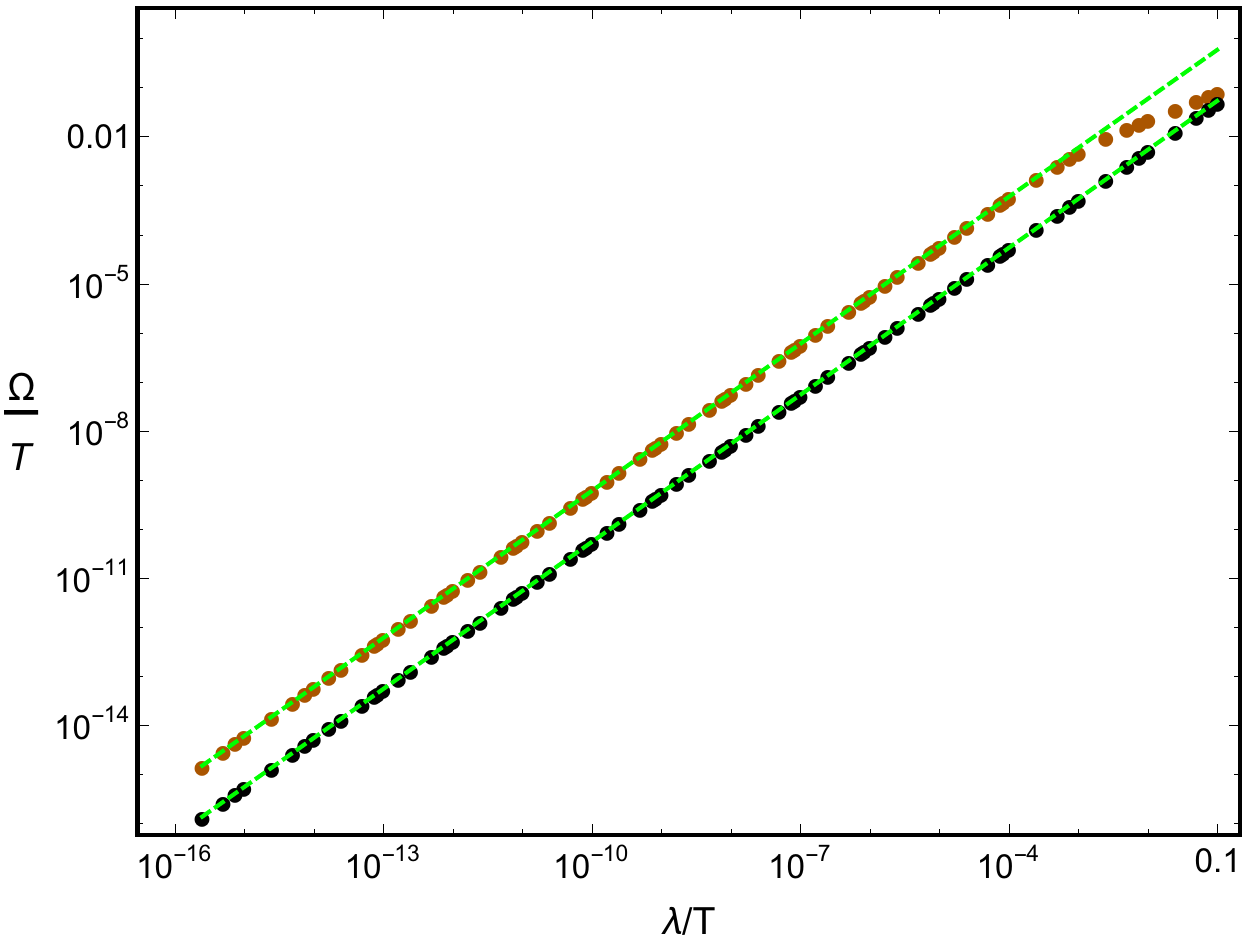}
    \caption{Colours indicate fixed $T/\mu$ -- light brown is $0.0582$ and black is $0.0434$. Dashed green lines mark a linear fit to the data. \textbf{Left:} The pinning frequency $\omega_0^2$ as a function of the dimensionless source $\lambda/T$. \textbf{Right: }The phase relaxation rate $\Omega$ as a function of the dimensionless source $\lambda/T$. At small explicit breaking, both quantities are linear in the explicit breaking parameter as emphasised by the fits shown with green dashed lines.}
    \label{fig3}
\end{figure}

The pinning frequency $\omega_0$ and phase relaxation rate $\Omega$, as functions of the dimensionless explicit breaking parameter $\lambda/T$, may be extracted directly from the zero-momentum quasi-normal mode data -- the results are shown in figure \ref{fig3}. The left panel illustrates that the pinning frequency $\omega_0^2$ vanishes (linearly) without the presence of explicit breaking; this is another realisation of the Gell-Mann-Oaks-Renner relation presented and discussed around equation \eqref{eq:GMOR}. In the right panel we see that the phase relaxation rate vanishes alongside the explicit breaking parameter, indicating that the appearance of $\Omega$ is dependent on the interplay between explicit and spontaneous symmetry breaking rather than being a generic feature of the model without explicit breaking.\footnote{The appearance of an effective phase relaxation has also been observed in holographic models of translational symmetry breaking \cite{Amoretti:2018tzw,Ammon:2019wci,Baggioli:2019abx,Andrade:2018gqk,Donos:2019txg,Donos:2019hpp,Andrade:2020hpu}.} Moreover, for small explicit breaking both quantities display a linear dependence on $\lambda$ -- this behaviour is typical of the pseudo-spontaneous regime and has been observed in several analogous situations \cite{Alberte:2017cch,Ammon:2019wci,Amoretti:2018tzw,Donos:2019tmo,Andrade:2020hpu,Andrade:2018gqk,Andrade:2017cnc}.

\begin{figure}
    \centering      \includegraphics[width=0.45\linewidth]{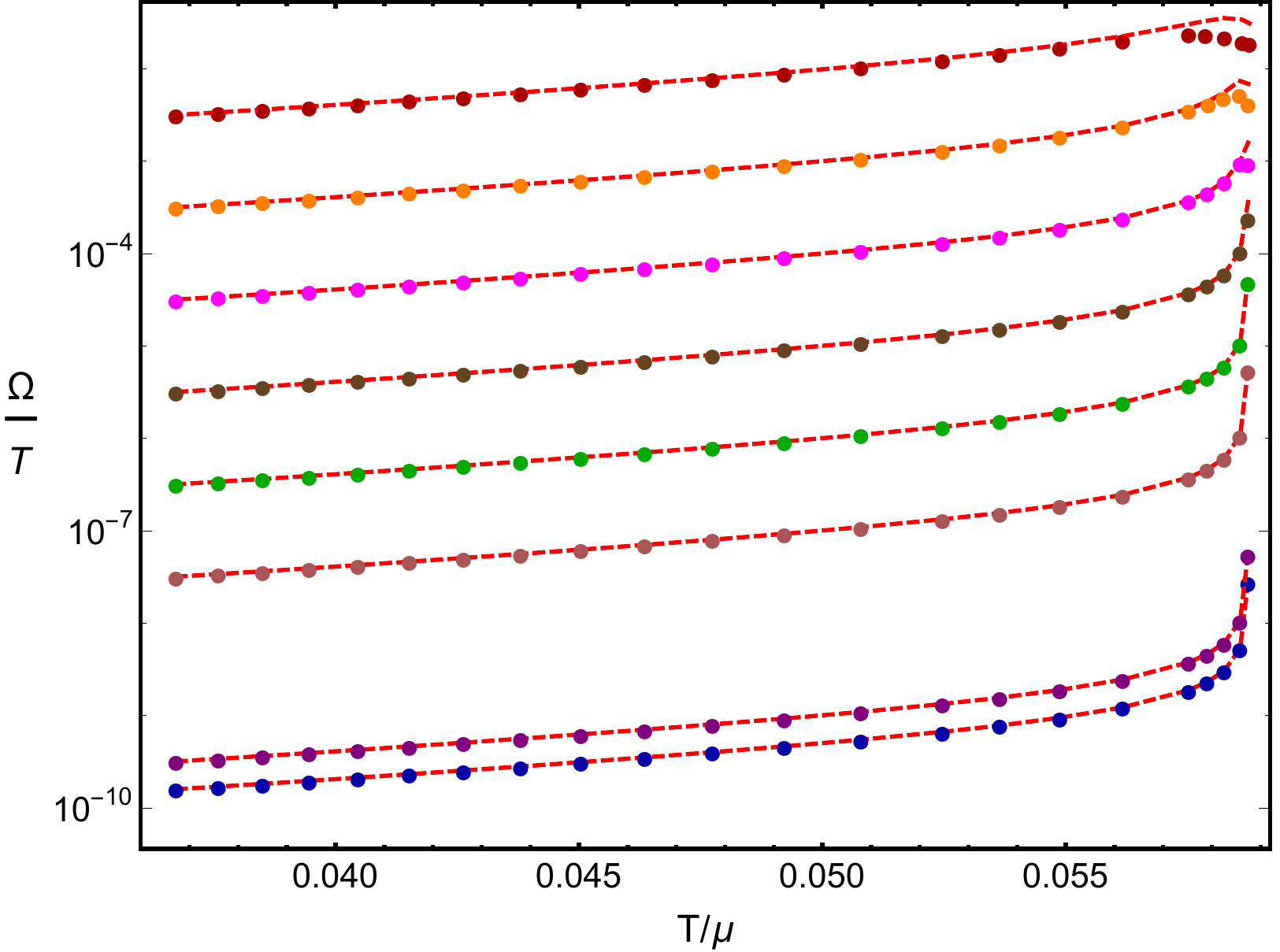}\quad
            \includegraphics[width=0.45\linewidth]{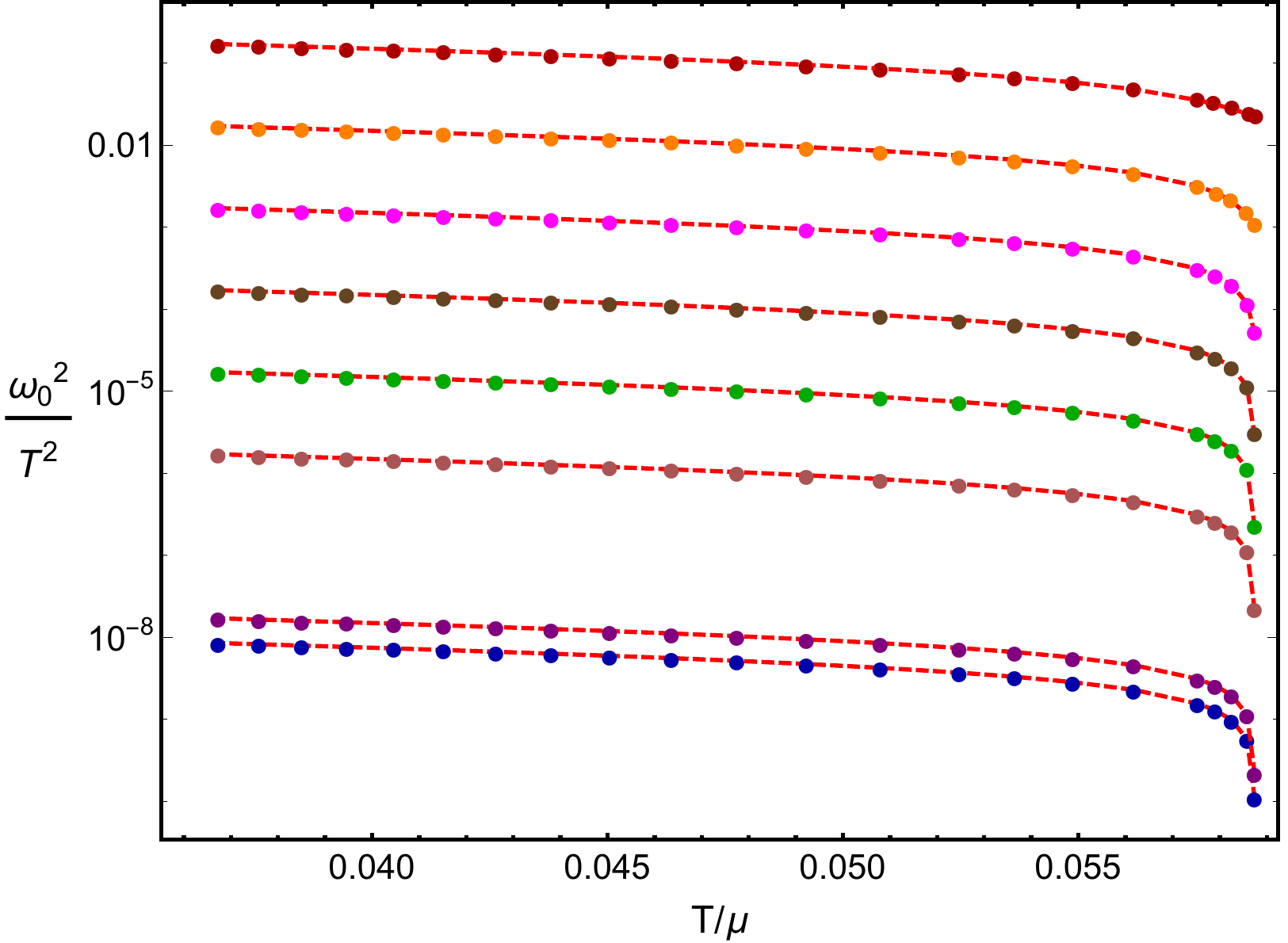}\newline
    \caption{The phase relaxation rate (\textbf{left}) and the pinning frequency (\textbf{right}) in function of the dimensionless temperature $T/\mu$ for different values of the explicit breaking parameter $\lambda/T$. Dots indicate quasi-normal mode data, while the dashed lines in the left panel are values of the Kubo formula \eqref{kubokubo} and in the right panel the dashed lines correspond to the definition  \eqref{pred1}. The colours indicate different values of the dimensionless source $\lambda/T$ -- blue is $5\times10^{-10}$, purple is $10^{-9}$, light brown is $10^{-7}$, green is $10^{-6}$, dark brown is $10^{-5}$, magenta is $10^{-4}$, orange is $10^{-3}$, and red is $10^{-2}$.}
    \label{fig:mm1}
\end{figure}

In figure \ref{fig:mm1} we show the values of the phase relaxation rate $\Omega$ and the pinning frequency $\omega_0^2$ as a function of $T/\mu$ for different values of the source $\lambda$; dots are quasi-normal mode data while the dashed lines indicate the Kubo formula \eqref{kubokubo} for $\Omega$ and the definition \eqref{pred1} for $\omega_0$. The agreement is good until large values of the explicit breaking parameter and when the critical temperature is approached -- this is a manifestation that in such a regime the explicit breaking is no longer a subleading effect to the spontaneous one and the physics is no longer governed by the pseudo-spontaneous approximation.


\subsection{Finite-momentum Excitations}
Moving into the realm of finite momentum excitations, we may use the holographic quasi-normal mode data to confirm our hydrodynamic framework presented in section \ref{sec0}. At vanishing charge relaxation $\Gamma$ the real and imaginary parts of the $k^2$-coefficient \eqref{fufu} are given by
\begin{subequations}\label{ImReD}
\begin{align}
    \mathrm{Re} \, D_\pm &= \frac12  \del{  \frac{ \sigma_0}{(\partial \rho_t / \partial \mu) } + \zeta_3  \frac{\rho_s}{\mu} }, \label{hydroform1}\\
    \mathrm{Im}\, D_\pm &= \pm \frac{1}{2} { \frac{ 2\rho_s - \zeta_3  \rho_s (\partial \rho_t / \partial \mu)\Omega + \sigma_0 \mu \Omega}{\mu\sqrt{4m(\partial \rho_t / \partial \mu) - (\partial \rho_t / \partial \mu)^2 \Omega^2 } } }.\label{hydroform2}
\end{align}
\end{subequations}
Note that the expressions \eqref{hydroform1} and \eqref{hydroform2} respectively constitute the imaginary and real part of the dispersion relations with finite momentum. 

\begin{figure}
    \centering      \includegraphics[width=0.45\linewidth]{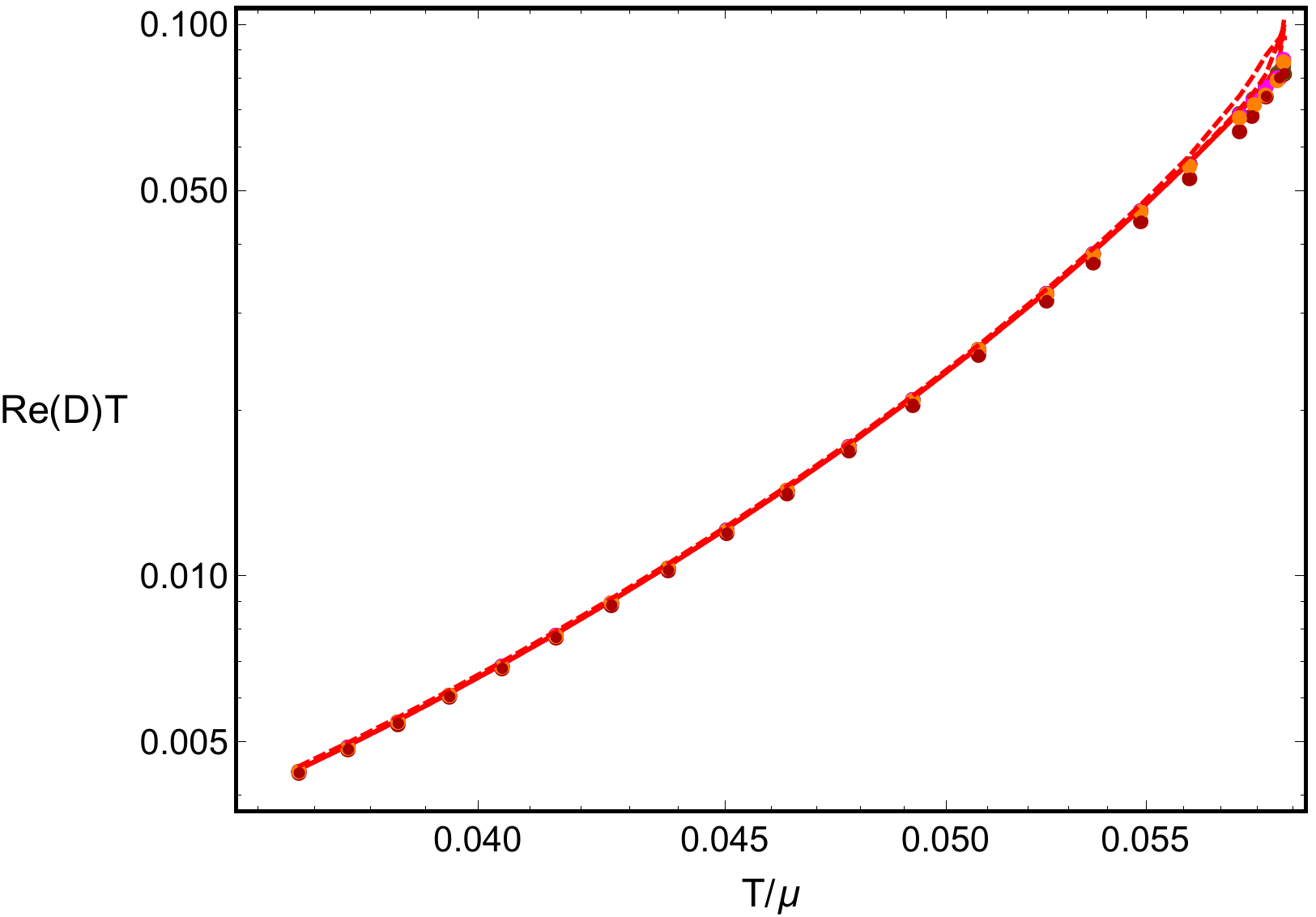}\quad \includegraphics[width=0.45\linewidth]{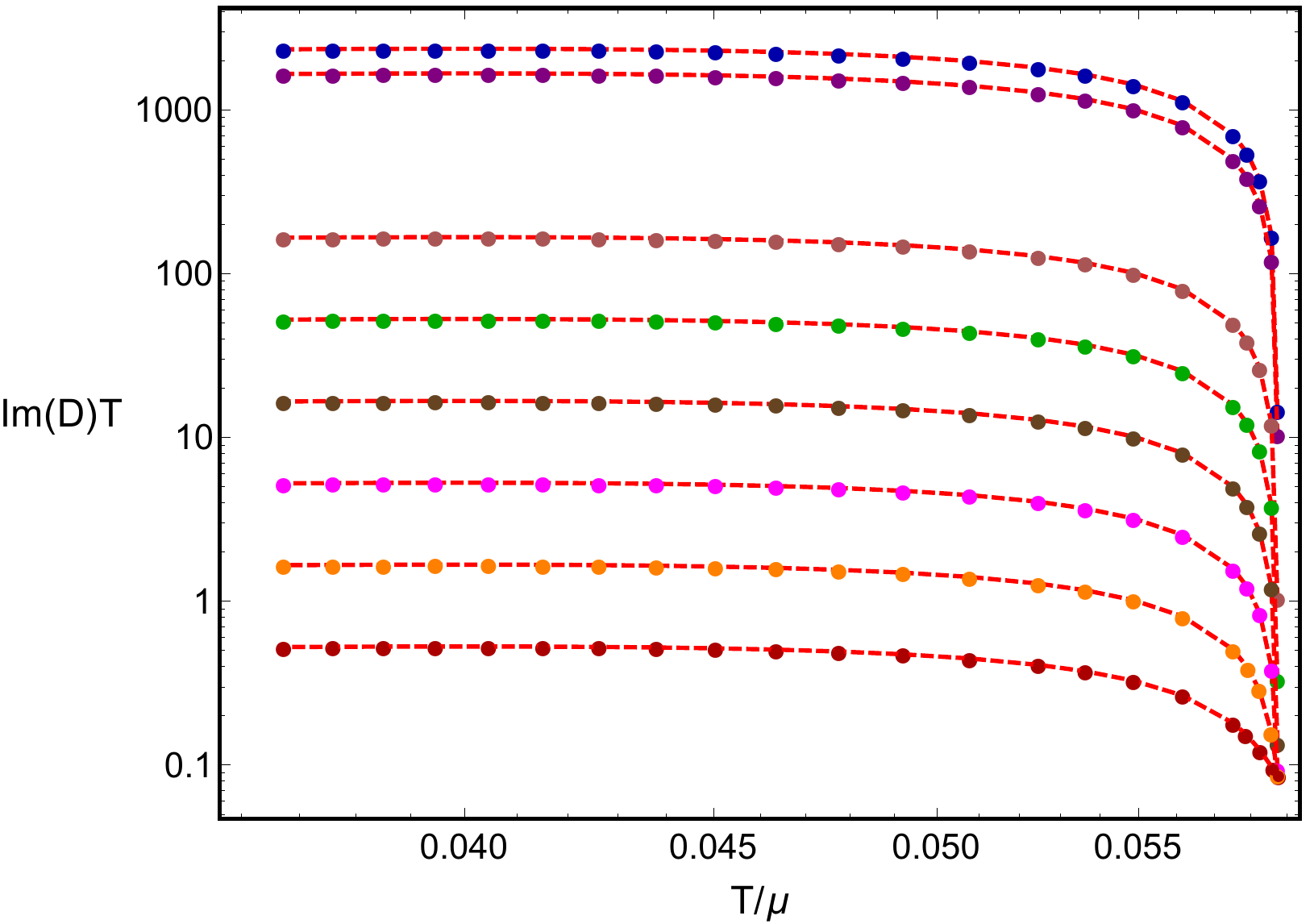} 
    
    \vspace{0.3cm}
    
    \includegraphics[width=0.6\linewidth]{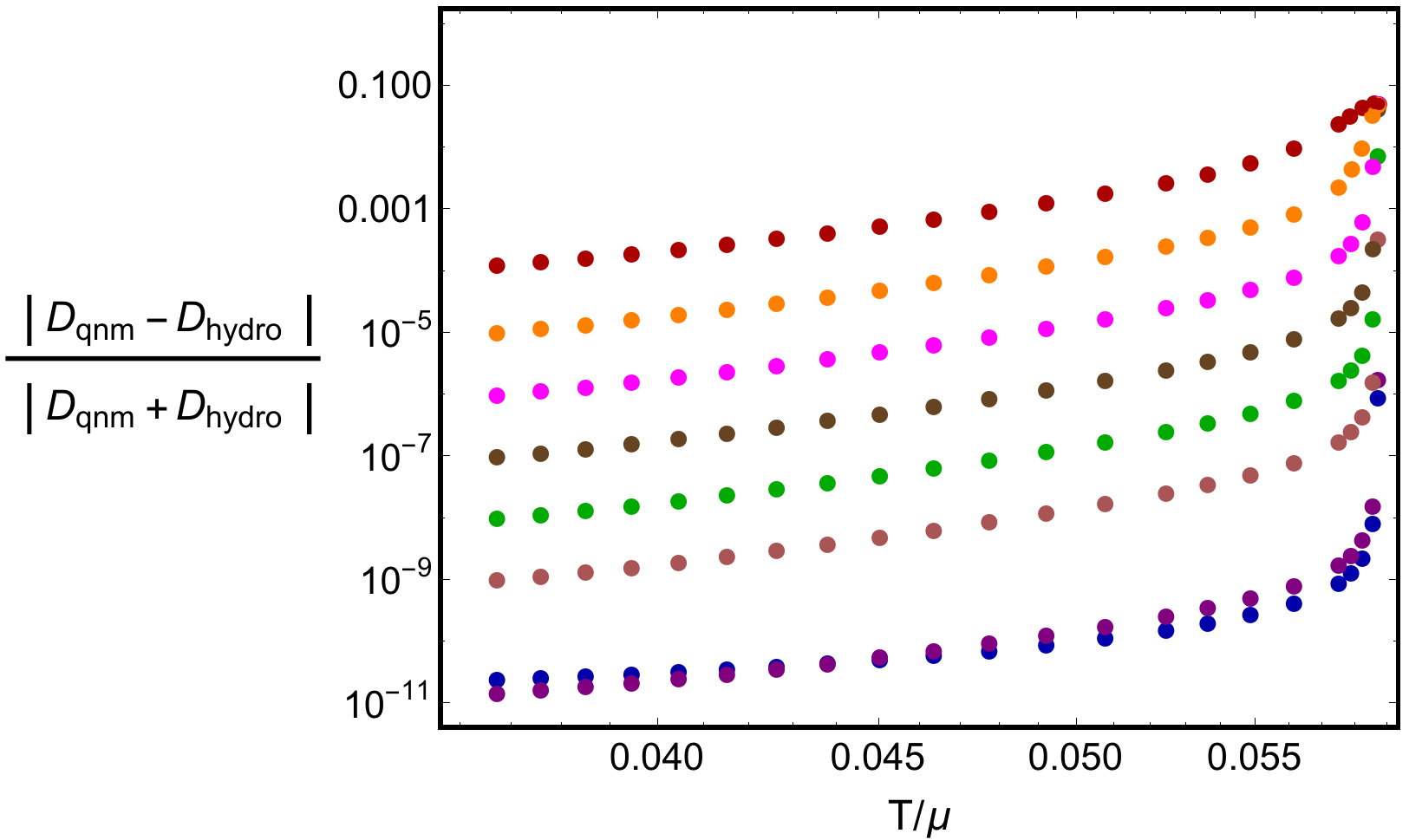}
    \caption{Comparison between the $k^2$-coefficients of the lowest quasi-normal modes (dots) versus the hydrodynamic formulae in \eqref{hydroform1} and \eqref{hydroform2} (lines), as a function of the dimensionless temperature $T/\mu$. The bottom panel shows the degree of agreement between the quasi-normal mode data and the hydrodynamic formulae. It shows increasing agreement by lowering the temperature, i.e. by making the order parameter for the spontaneous breaking larger compared to the explicit breaking scale $\lambda/T$. The colours indicate different values of the dimensionless source $\lambda/T$ -- blue is $5\times10^{-10}$, purple is $10^{-9}$, light brown is $10^{-7}$, green is $10^{-6}$, dark brown is $10^{-5}$, magenta is $10^{-4}$, orange is $10^{-3}$, and red is $10^{-2}$.}
    \label{fig:mm2}
\end{figure}

The hydrodynamic formulae \eqref{ImReD} are plotted alongside the holographic data in the top panel of figure \ref{fig:mm2}. The agreement is good and tested for several, small values of the dimensionless explicit breaking parameter $\lambda/T$; we do not expect this match to hold away from the pseudo-spontaneous regime. To gain further insight into the pseudo-spontaneous approximation we in the bottom panel of figure \ref{fig:mm2} compare the hydrodynamic predictions in \eqref{ImReD} to numerical data while varying $T/\mu$. A smaller value of $T/\mu$ corresponds to a larger value of the condensate and therefore the pseudo-spontaneous limit provides a reasonable approximation -- in fact, the agreement gets better when moving to low temperatures.\footnote{Unfortunately, due to the probe limit approximation, we cannot extend our analysis to very low temperatures to see when the approximation fails.}


\subsection{On the Universality of the Effective Phase Relaxation}
\label{uniuni}
We now turn our attention to an aspect of the phase relaxation which is often referred to as universal. It has been observed in a number of works, in various contexts, that an effective phase relaxation induced by pseudo-spontaneous symmetry breaking is related to the pinning frequency as well as the mass and diffusivity of the pseudo-Goldstone mode \cite{Amoretti:2018tzw,Ammon:2019wci,Baggioli:2019abx,Andrade:2018gqk,Donos:2019txg,Donos:2019hpp,Andrade:2020hpu,Baggioli:2020nay,Baggioli:2020haa,Amoretti:2021fch,Grossi:2020ezz}.\footnote{The first observation of a potentially universal behaviour appeared in \cite{Amoretti:2018tzw}, for a holographic homogeneous Q-lattice model which breaks translational invariance. It has since been confirmed numerically in several holographic models with translational symmetry breaking \cite{Ammon:2019wci,Baggioli:2019abx,Andrade:2018gqk,Donos:2019txg,Donos:2019hpp,Andrade:2020hpu}. A first derivation using effective field theory methods was proposed in \cite{Baggioli:2020nay} and later completed using Keldysh-Schwinger techniques in \cite{Baggioli:2020haa}. The universal relation in the presence of a finite magnetic field has also been discussed in \cite{Amoretti:2021fch}, and in the context of the chiral limit \cite{Grossi:2020ezz}.} The general form of the relation may be expressed in several ways -- one of them is \cite{Baggioli:2020haa}
\begin{equation}
    \Omega=\omega_0^2\,\frac{D_\xi}{v_s^2}
\end{equation}
where $D_\xi$ and $v_s$ respectively denote the Goldstone diffusivity and speed of sound in the purely spontaneous phase.\footnote{The diffusivity $D_\xi$ does in general not correspond to a diffusive mode unless the Goldstone dynamics decouples, in which case it can be directly extracted from the Josephson relation \eqref{josephsonDeriv}.} We hence conclude that 
\begin{equation}
    \Omega\,=\,\omega_0^2\,\zeta_3\,\chi_{\rho\rho}\label{uni1}
\end{equation}
for the case of pseudo-spontaneous breaking of $U(1)$ symmetry, with $D_\xi = \zeta_3 \rho_s/\mu$. The quantities entering the above equation are defined in the hydrodynamic framework of section \ref{sec0}.\footnote{In appendix \ref{uniapp} we derive an analogous expression using the holographic dispersion relations of \cite{Donos:2021pkk}.} Importantly, while $\Omega$ and $\omega_0^2$ are linear in the explicit breaking scale for sufficiently small breaking, $\zeta_3$, $\chi_{\rho\rho}$ are of order $\mathcal{O}\left(\lambda^0\right)$ and finite even in absence of explicit symmetry breaking. In fact, at leading order in the explicit breaking the $\zeta_3$ entering in the above relation is that of the purely spontaneous phase. Implementing the appropriate mappings, the expression \eqref{uni1} is in agreement with the analogous expressions in the case of translational symmetry breaking, see for instance \cite{Amoretti:2018tzw}.

\begin{figure}
    \centering
     \includegraphics[width=0.65\linewidth]{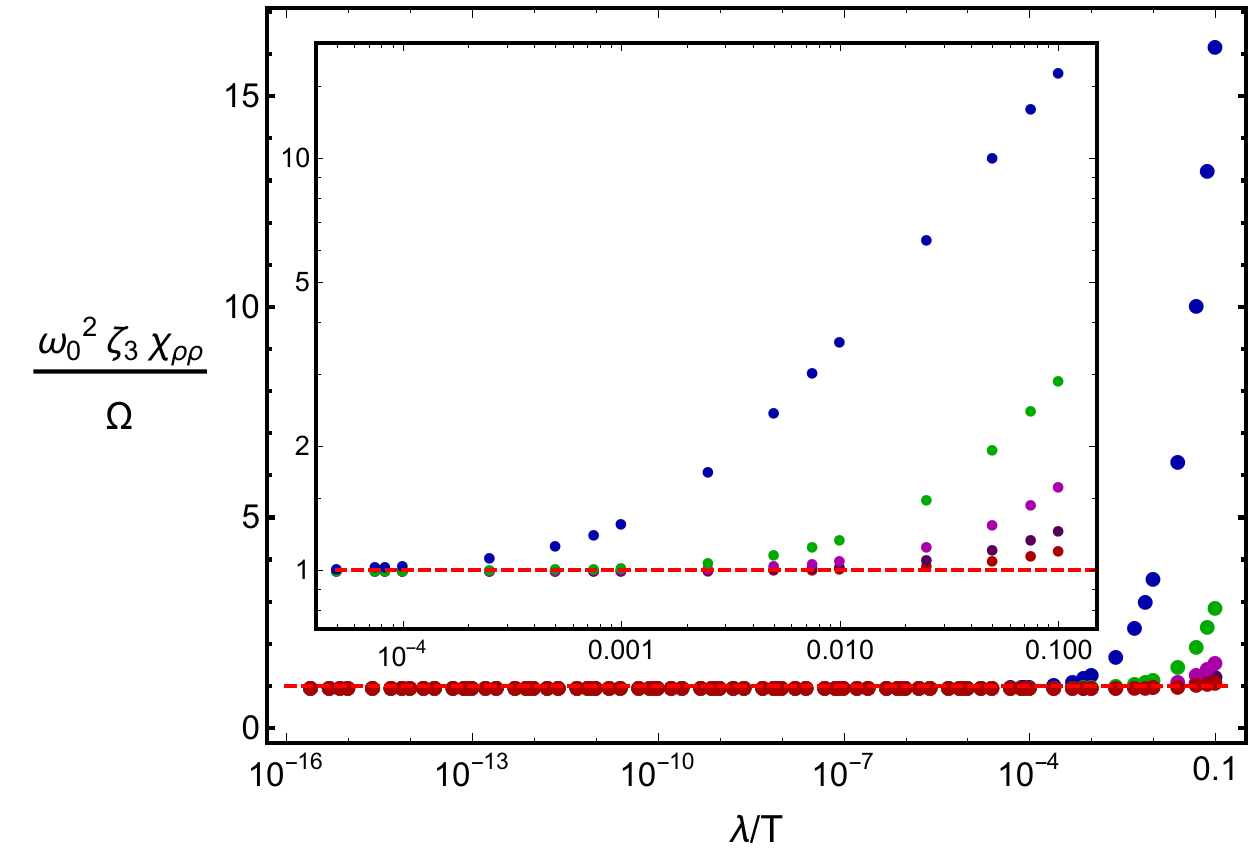}
    \caption{The dimensionless ratio $\omega_0^2 \zeta_3 \chi_{\rho\rho}/ \Omega$ as a function of the dimensionless source $\lambda/T$. At small explicit breaking, the value of this ratio reaches unity as expected from the universal relation in equation \eqref{uni1}. Different colours indicate different values of $T/\mu =\{0.0582, 0.0555, 0.0519, 0.0477, 0.0434\}$, starting from blue. We hence see a better agreement for larger chemical potentials which correspond to a larger spontaneous symmetry breaking scale and therefore a better validity for the pseudo-spontaneous approximation.}
    \label{figUni}
\end{figure}

The task at hand is to test if equation \eqref{uni1} holds for our version of the holographic model \eqref{eq:action}. The ratio $\omega_0^2 \zeta_3 \chi_{\rho\rho}/\Omega$ is plotted in figure \ref{figUni}. We see that the ratio tends to unity in the regime of small explicit breaking, from which we may conclude that the relation \eqref{uni1} holds. This constitutes additional evidence for the universal behaviour of the effective phase relaxation, at least for small explicit breaking. In fact, $\Omega \leq \omega_0^2 \zeta_3 \chi_{\rho\rho}$ for all values of $\lambda$, indicating that $\omega_0^2 \zeta_3 \chi_{\rho\rho}$ acts as an upper bound. For large values of the explicit breaking parameter this relation ceases to be valid since the pseudo-Goldstone perspective must be abandoned. From figure \ref{figUni}, one sees that the validity of the relation \eqref{uni1} corresponds with that of the pseudo-spontaneous limit -- in fact, by decreasing the temperature and going towards larger values of the condensate the relation holds for a larger range of $\lambda$.

\subsection{The Electric Conductivity}
Another noteworthy result which originates from our hydrodynamic theory is the behaviour of the electrical conductivity; see equation \eqref{cc} for the hydrodynamic AC conductivity. Due to the effects of pseudo-spontaneous symmetry breaking the DC conductivity becomes finite but without displaying a low frequency Drude-like form, in contrast to the behaviour of superfluids with phase relaxation caused by vortices \cite{PhysRev.140.A1197,Halperin1979}.

\begin{figure}
    \centering
     \includegraphics[width=0.45\linewidth]{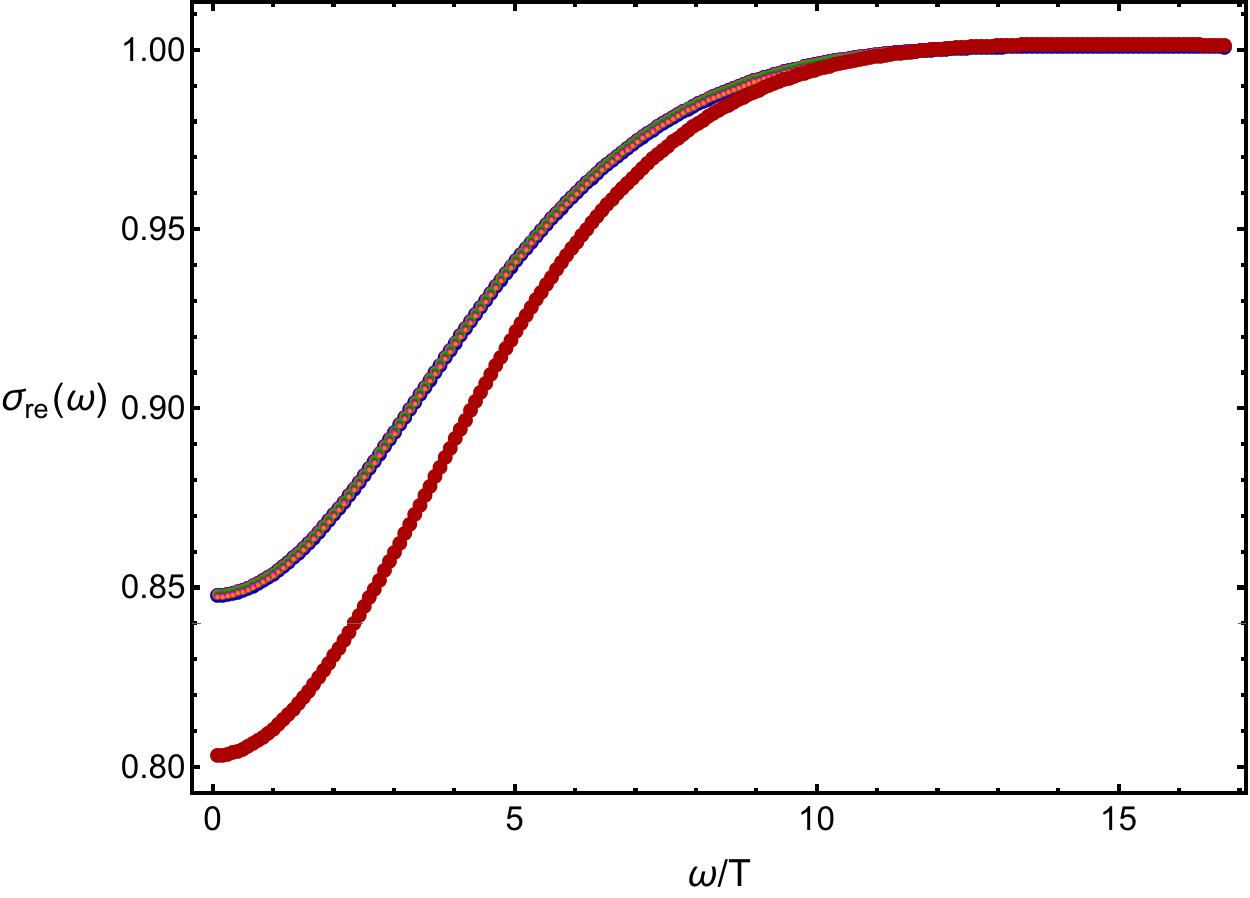}\quad
           \includegraphics[width=0.45\linewidth]{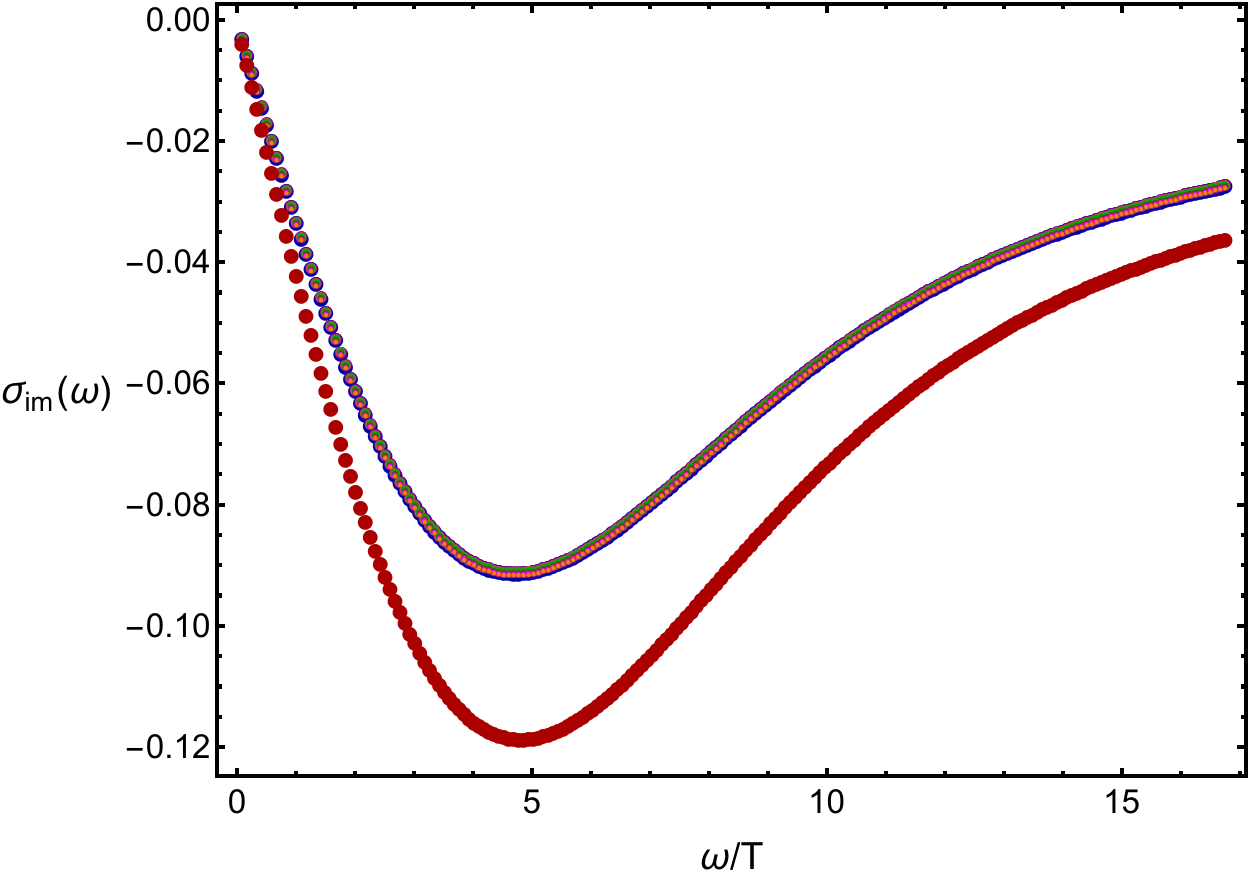}
   \caption{Real (\textbf{left}) and imaginary (\textbf{right}) parts of the AC conductivity at fixed $T/\mu = 0.0579$, as a function of the dimensionless frequency $\omega/T$, for different values of the dimensionless source $\lambda/T$. The colours indicate different values of the dimensionless source $\lambda/T$ -- blue is $5\times0^{-10}$, purple is $10^{-9}$, light brown is $10^{-7}$, green is $10^{-6}$, dark brown is $10^{-5}$, magenta is $10^{-4}$, orange is $10^{-3}$, and red is $10^{-2}$. The data from most cases are on top of each other.}
    \label{figCondictivity}
\end{figure}

We plot the frequency dependent conductivity for different values of the dimensionless source $\lambda/T$ in figure \ref{figCondictivity}. The numerical data follow the theoretical prediction
\begin{equation}\label{conductivity2}
    \sigma(\omega) =  \sigma_0 \,+\mathcal{O}(\omega^2)\, \,.
\end{equation}
Some observations are in order. First, the incoherent conductivity is decreased by the presence of the finite dimensionless source $\lambda/T$ and a gap opens in the real part of the conductivity upon increasing $\lambda/T$. It may be possible to obtain a generalised formula for the incoherent conductivity in terms of horizon data which could explain this behaviour analytically, but we leave this task for future work. Second, the real part of the conductivity grows with a quadratic scaling $(\mathrm{Re}[\sigma]-\sigma_0) \sim \omega^2$ at low frequency, which is not captured by the lowest order hydrodynamic expression \eqref{conductivity2}. Third, the imaginary part of the conductivity goes to zero linearly with the frequency implying that the superfluid pole is now removed. Finally, and most importantly, the value of the DC conductivity not only becomes finite but it is given exactly by the incoherent contribution $\sigma_0$.

From a physical perspective the fundamental result is that the optical conductivity does not display an infinite DC conductivity nor a Drude-like structure, in contrast to the scenario where phase relaxation is induced by defects \cite{PhysRevB.94.054502}. The reason for this may be understood as follows. In the case of defects the superfluid sound mode acquires a finite lifetime $\Omega$ but not a real gap $\omega_0$. As such, the dynamics of the conductivity at low frequency is the same as that of the Drude model, where the width of the Drude peak is given by the phase relaxation rate $\Omega$. In contrast, for our version of the model \eqref{eq:action} the lowest-lying mode acquires a finite mass gap as well, given by the pinning frequency $\omega_0$.

As a final comment, in view of possible condensed matter applications, our findings may be crucial to understand whether phase relaxation, and which type of it, plays any role in the physics of the so-called 2D failed superconductors \cite{Phillips_2003}.

\section{Holographic Superfluid with a Massive Gauge Field}\label{sec:four}
In this section we study an implementation of pseudo-spontaneous breaking of a $U(1)$ symmetry through holography which is alternative to that of section \ref{sec1}. We consider a modification of the holographic superfluid \cite{Hartnoll:2008vx} in which the gauge vector dual to the $U(1)$ current is massive, thus making the current anomalous and explicitly breaking the symmetry. This model was considered in \cite{Jimenez-Alba:2015awa,Jimenez-Alba:2014iia} to study the explicit
breaking of a global $U(1)$ symmetry.
We will again restrict the analysis to the probe limit and consider a field theory in $(2+1)$-dimensions.\footnote{Adding backreaction to the model \eqref{eq:action2}  results in a dual Lifshitz field theory. However, if the mass of the bulk gauge field is sufficiently small the dual field theory may also be treated as a relativistic CFT deformed by the timelike component of its $U(1)$ current \cite{Korovin:2013bua,Korovin:2013nha, Taylor:2015glc}.} As we will see, the phenomenology and features of this model are markedly different from the one examined in the previous section, thus providing a complementing perspective.


\label{sec2}
\subsection{The Holographic Model}
We consider the following bulk action
\begin{equation}
\label{eq:action2}
S\,=\,\int \dd^{4}x\, \sqrt{-g}\left[-\frac{1}{4}F_{\mu\nu}F^{\mu\nu}-\frac{M_A^2}{2}\,\left(A_\mu-\partial_\mu \theta\right) \left(A^\mu-\partial^\mu \theta\right)-|D\psi|^2-M^2|\psi|^2\right]\,,
\end{equation}
where $M_A$ is the bulk mass for the gauge field $A_\mu$. The mass term breaks the local $U(1)$ symmetry $ A_\mu \mapsto A_\mu +\partial_\mu \lambda_g$; however, the St\"uckelberg field $\theta$ transforms as $\theta \mapsto \theta - \lambda_g$ under the local $U(1)$ symmetry, restoring the gauge invariance of the action \eqref{eq:action2}.\footnote{Notice the analogy with holographic models of broken translations where the graviton becomes massive \cite{Alberte:2015isw}.} The background spacetime geometry is given by the line element \eqref{eq:metric}. 

The presence of a finite bulk mass modifies the asymptotics of the bulk gauge field into
\begin{equation}
    A_\mu(z)\sim \,A_\mu^{(l)}\,z^{-\Delta_A}(1+\dots)+A_\mu^{(s)}\,z^{1+\Delta_A}(1+\dots)\qquad \text{with}\qquad \Delta_A= \frac{1}{2} \left( -1 + \sqrt{1+ 4 \, M_A^2}  \right). \label{eq11}
\end{equation}
The expectation value of the $U(1)$ current, $\braket{J_\mu}$, is related to $A_\mu^{(s)} - \partial_\mu \theta^{(s)}$ and thus the scaling dimension is $\left[J^\mu\right]=2+\Delta_A$, where $\Delta_A$ is the anomalous scaling dimension indicating that the $U(1)$ current is no longer conserved \cite{Jimenez-Alba:2014iia,Jimenez-Alba:2015awa}. Increasing $M_A^2$ makes the explicit breaking larger and hence the scaling dimension more anomalous. We again define a quantity $\mu$ to be given by the leading asymptotic contribution of $A_t$ as $\mu = A_t^{(l)}$ -- note that $\mu$ should be viewed as a coupling in the Lagrangian of the dual boundary field theory and \textit{not} as the chemical potential.\footnote{We consider static configurations in which $\partial_t \theta=0$.} The equations of motion for the gauge field constrain its temporal component to vanish at the horizon. Moreover, the chemical potential, defined as the integrated radial electric flux in the bulk, is infinite \cite{Jimenez-Alba:2014iia} -- this does not give rise to any instabilities in the quasi-normal mode spectrum, at least in the probe limit.

The scalar field $\psi$ is charged under the $U(1)$ symmetry, as in section \ref{sec1}, and we will choose its mass to be $M^2=-2$. In contrast to the previous section we will fix the source of the charged scalar field, $\lambda$, to zero. In terms of the boundary expansion \eqref{eq:bdryexp}, this gives rise to  the boundary condition $\psi_i^{(l)}=0$. Nevertheless, for small enough temperatures or large values of $\mu$ we find solutions with a non-zero vacuum expectation value for the dual operator $\mathcal{O}$, i.e. with $\psi^{(s)}\neq 0$; thus, when the mass of the bulk gauge field is `small' the $U(1)$ symmetry will be broken pseudo-spontaneously as in the previous toy model.

Next, we turn to the Ward identity for the current. For holographic dualities involving four-dimensional boundary gauge theories the Stückelberg field $\theta$ is dual to the topological charge density of the dynamical gauge fields  \cite{Liu:1998ty,Klebanov:2002gr,Iatrakis:2015fma,Bigazzi:2018ulg}. The topological charge density in field theory appears in the non-conservation of the current $J^\mu$ \cite{Jimenez-Alba:2014iia,Iatrakis:2015fma}, which hence should be considered to be of axial nature. Fluctuations of the topological charge are related to (fluctuations of) the axial charge density $\braket{J^t}$, as argued in \cite{Iatrakis:2014dka} from field theoretic and holographic perspective -- see also \cite{PhysRevD.78.074033} for earlier work. Following this line of argumentation we arrive at the non-conservation equation\footnote{This relation may only be valid for fluctuations, which could explain the stability of the equilibrium state mentioned above.}
\begin{equation}\label{bb2}
    \partial_\mu \braket{J^\mu}=-\Gamma \braket{J^t} \, .
\end{equation}
For the moment we take \eqref{bb2} as a phenomenological equation also for the three-dimensional field theory dual to the gravitational theory \eqref{eq:action2}; this treatment is in agreement with the numerical results below.

Comparing the non-conservation equation \eqref{bb2} with equation \eqref{chargeNonCons} in the hydrodynamic framework we conclude that the parameter $m$, and hence the pinning frequency $\omega_0$, is not present for the model \eqref{eq:action2}. The dynamics of the quasi-normal modes will confirm that this model does not display phase relaxation.

\begin{figure}
    \centering
    \includegraphics[width=0.6\linewidth]{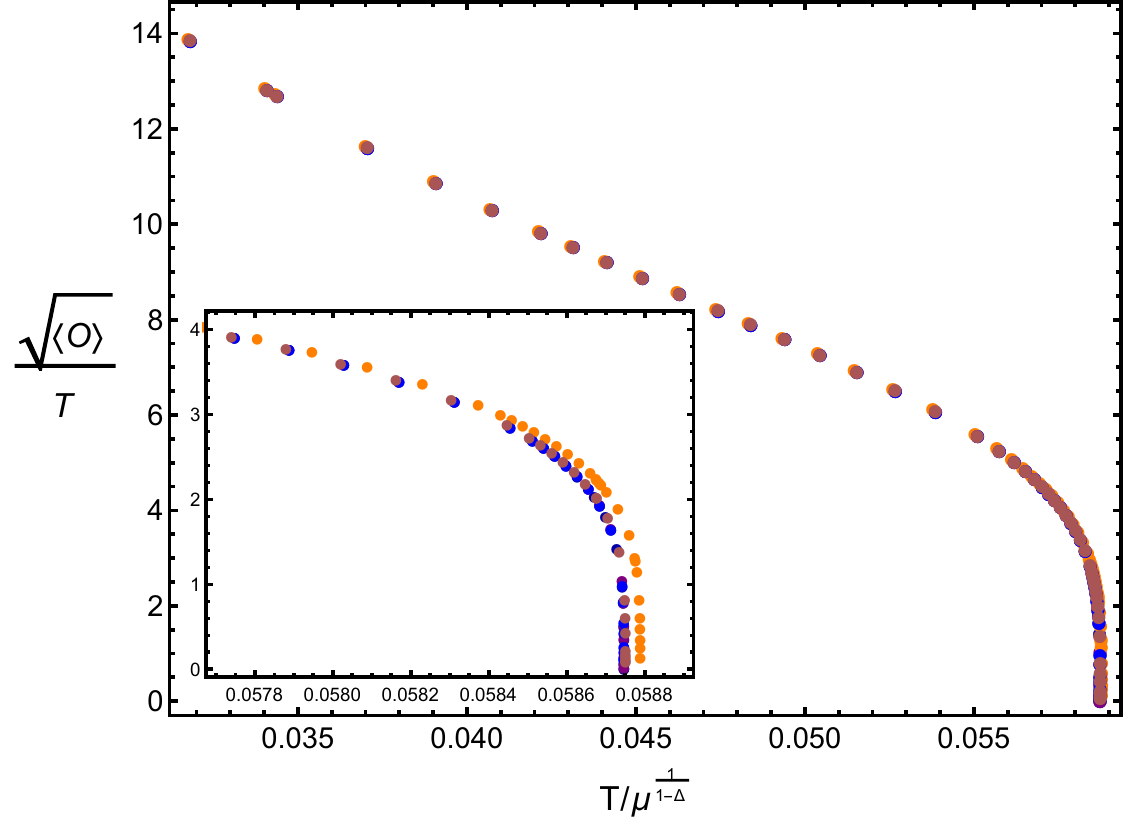}
    \caption{The behaviour of the dimensionless scalar condensate, as a function of the dimensionless temperature $T/\mu^{1/(1-\Delta_A)}$, for different amounts of explicit breaking. Colours indicate different values of the breaking parameter $M^2_A$ -- dark blue is $10^{-8}$, blue is $10^{-6}$, purple is $10^{-5}$, light brown is $10^{-4}$, and orange is $10^{-3}$. The condensate and the critical temperature grow with the breaking but the phase transition remains sharp.}
    \label{fig:mass1}
\end{figure}
Before analysing the dynamics of the model \eqref{eq:action2} we briefly comment on the thermodynamics of its pseudo-spontaneously broken phase. In figure \ref{fig:mass1} we show numerical data for the dimensionless scalar condensate at different values of the normalised bulk mass $M_A$. We observe how a finite, small bulk mass $M_A$ increases the value of the condensate and the critical temperature of the system but the phase transition remains sharp (in contrast to the model in section~\ref{sec1}). An intuitive argument for this behaviour is that, in this model, the Goldstone mode is not relaxing and thus there is still a sharp and clear definition of order.


\subsection{Quasi-normal Modes and Hydrodynamics}
\begin{figure}
    \centering
    \includegraphics[width=0.48\linewidth]{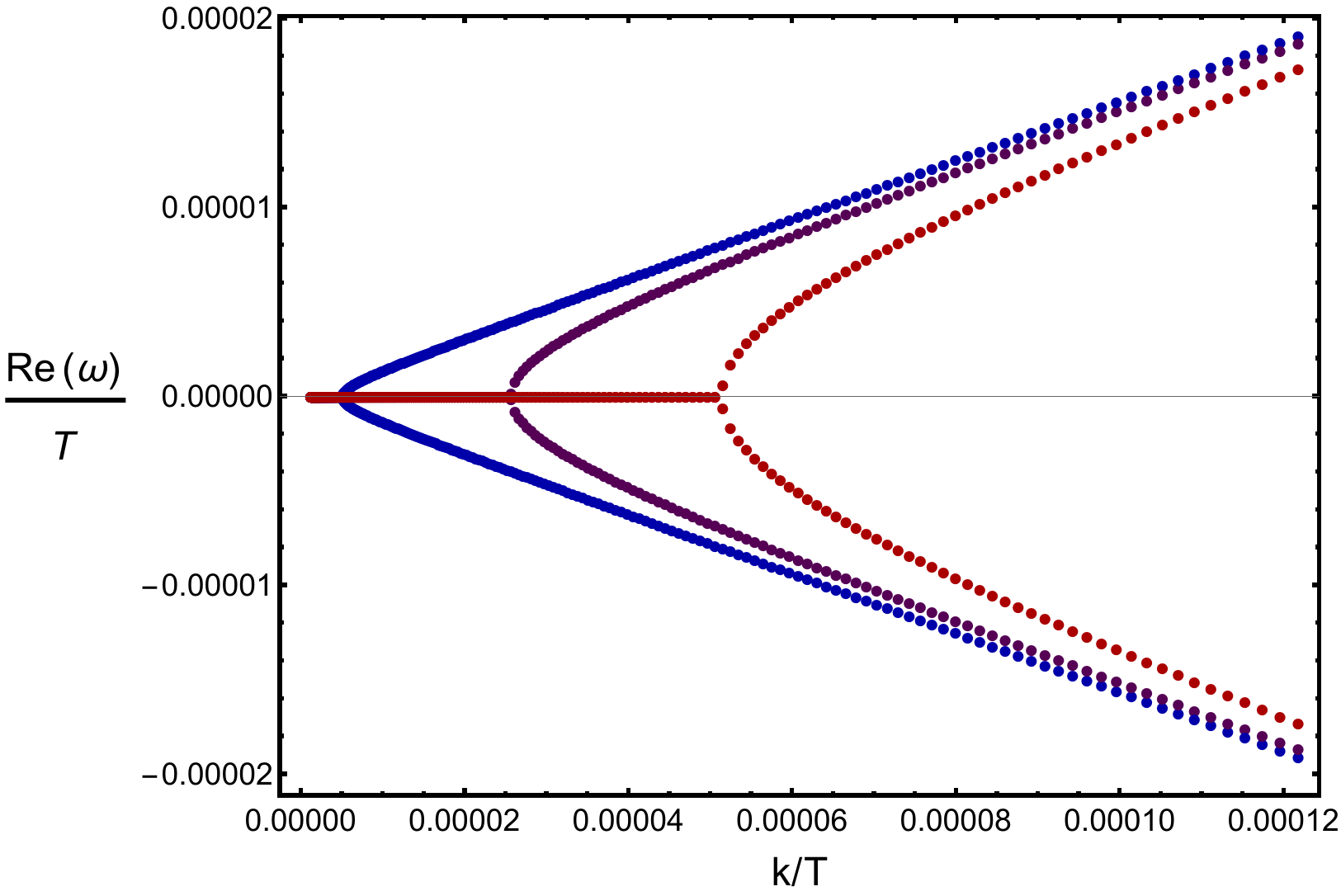}
    \includegraphics[width=0.48\linewidth]{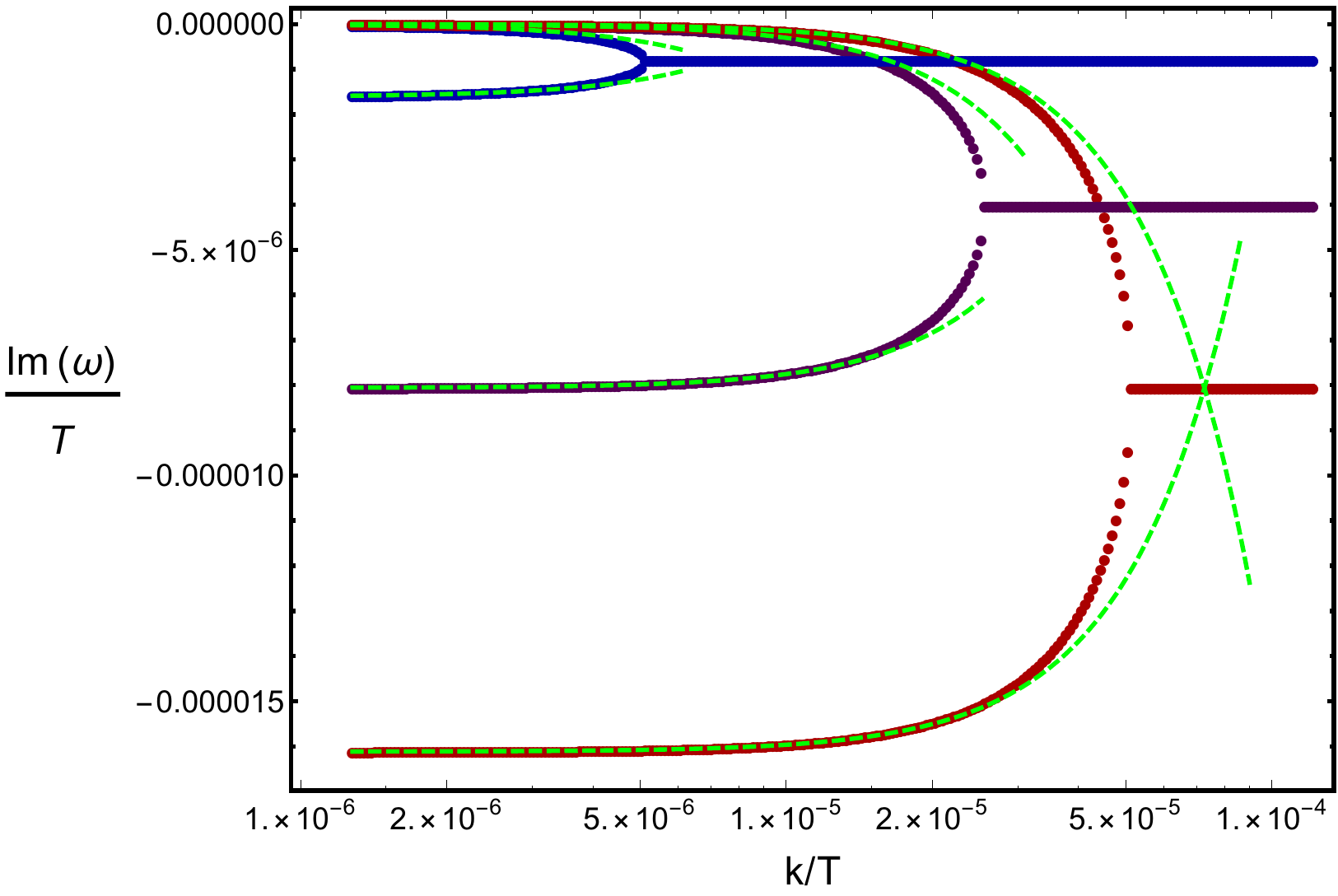}
    \caption{Low-lying quasi-normal modes as a function of the dimensionless momentum $k/T$ (dots) compared to hydrodynamic predictions \eqref{didi0} and \eqref{didi} using transport coefficient and susceptibility data extracted from the purely spontaneous phase (dashed lines, right panel). Colours indicate different values of the breaking parameter $M^2_A$ -- blue is $10^{-6}$, purple is $10^{-5}$, and red is $5 \times 10^{-5}$. Values are shown for fixed dimensionless temperature  $T/\mu^{1/(1-\Delta_A)}=0.0582$.
    }
    \label{fig7a}
\end{figure}

\begin{figure}
    \centering
    \includegraphics[width=0.6\linewidth]{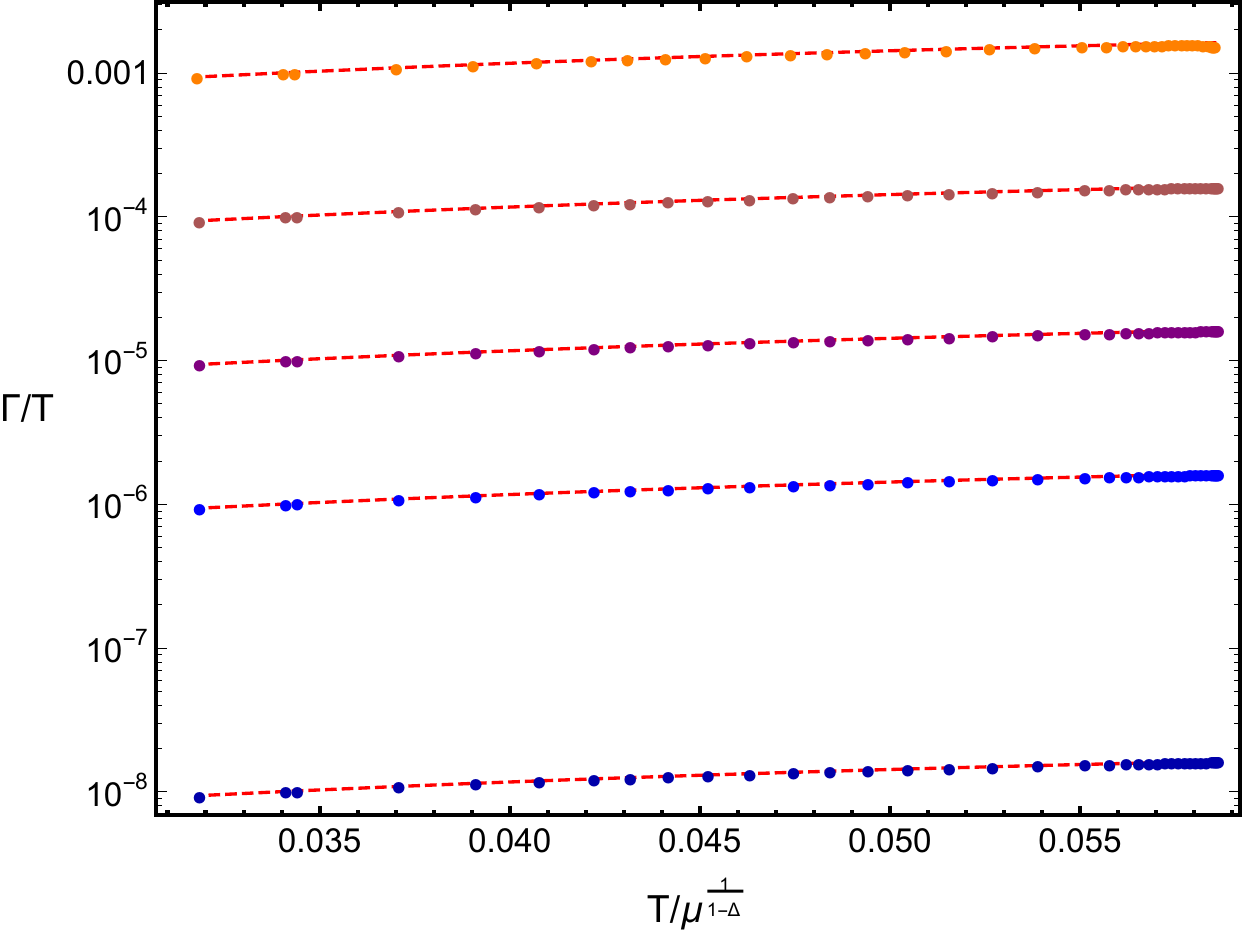}
    \caption{The dimensionless charge relaxation rate $\Gamma/T$ extracted from quasi-normal modes (dots) compared to the formula \eqref{rr}(dashed lines), as functions of the dimensionless temperature $T/\mu^{1/(1-\Delta_A)}$. Colours indicate different values of the breaking parameter $M^2_A$ -- dark blue is $10^{-8}$, blue is $10^{-6}$, purple is $10^{-5}$, light brown is $10^{-4}$, and orange is $10^{-3}$. The agreement is remarkable.}
    \label{fig6}
\end{figure}

By comparing the structure of the quasi-normal modes of this system (see figure \ref{fig7a}) to the hydrodynamic relation for the gap in equation \eqref{gap}, and taking the non-conservation of the current \eqref{bb2} into account, we conclude that phase relaxation does not appear in this model. In figure \ref{fig6} we plot the relaxation rate $\Gamma$ as a function of the mass of the bulk gauge field, $M_A$; for small values of $M_A$ we find
\begin{equation}
    \Gamma\,= \frac{M_A^2}{\chi_{\rho\rho}}+\dots, \label{rr}
\end{equation}
where the ellipsis indicate terms of higher order in $M_A$ and $\chi_{\rho\rho}$ is given by the purely spontaneous state. The result \eqref{rr} is reminiscent of the Drude relaxation rate in massive gravity models \cite{Davison:2013jba,Baggioli:2018vfc,Vegh:2013sk,Baggioli:2014roa}. The same behaviour is observed in the context of axial current dissipation in \cite{Jimenez-Alba:2015awa,Jimenez-Alba:2014iia}. It may be possible to perturbatively derive equation \eqref{rr} following the methods used in \cite{Davison:2013jba} -- we leave this task for future work.

\begin{figure}
    \centering
    \includegraphics[width=0.45\linewidth]{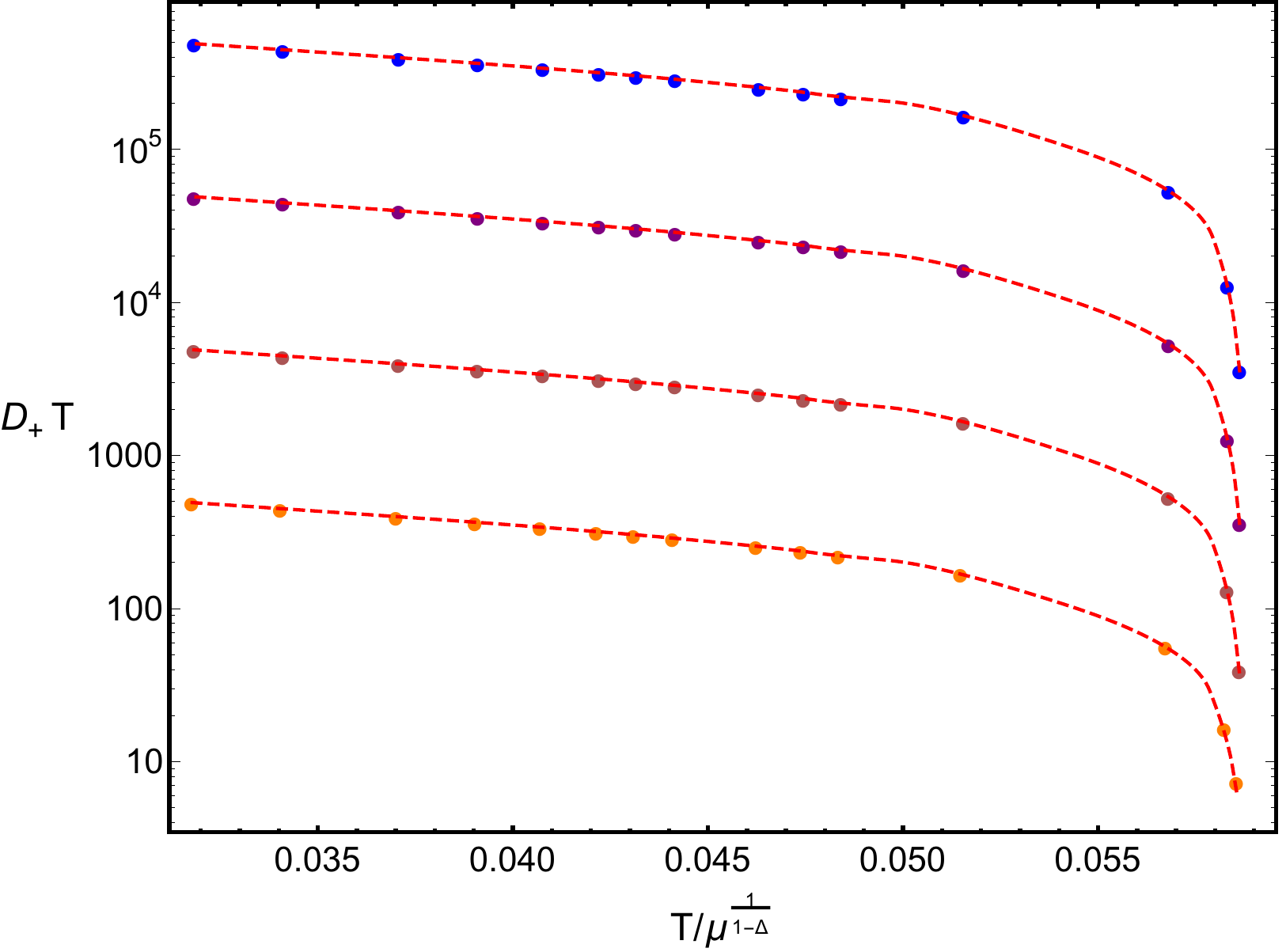}\qquad \includegraphics[width=0.45\linewidth]{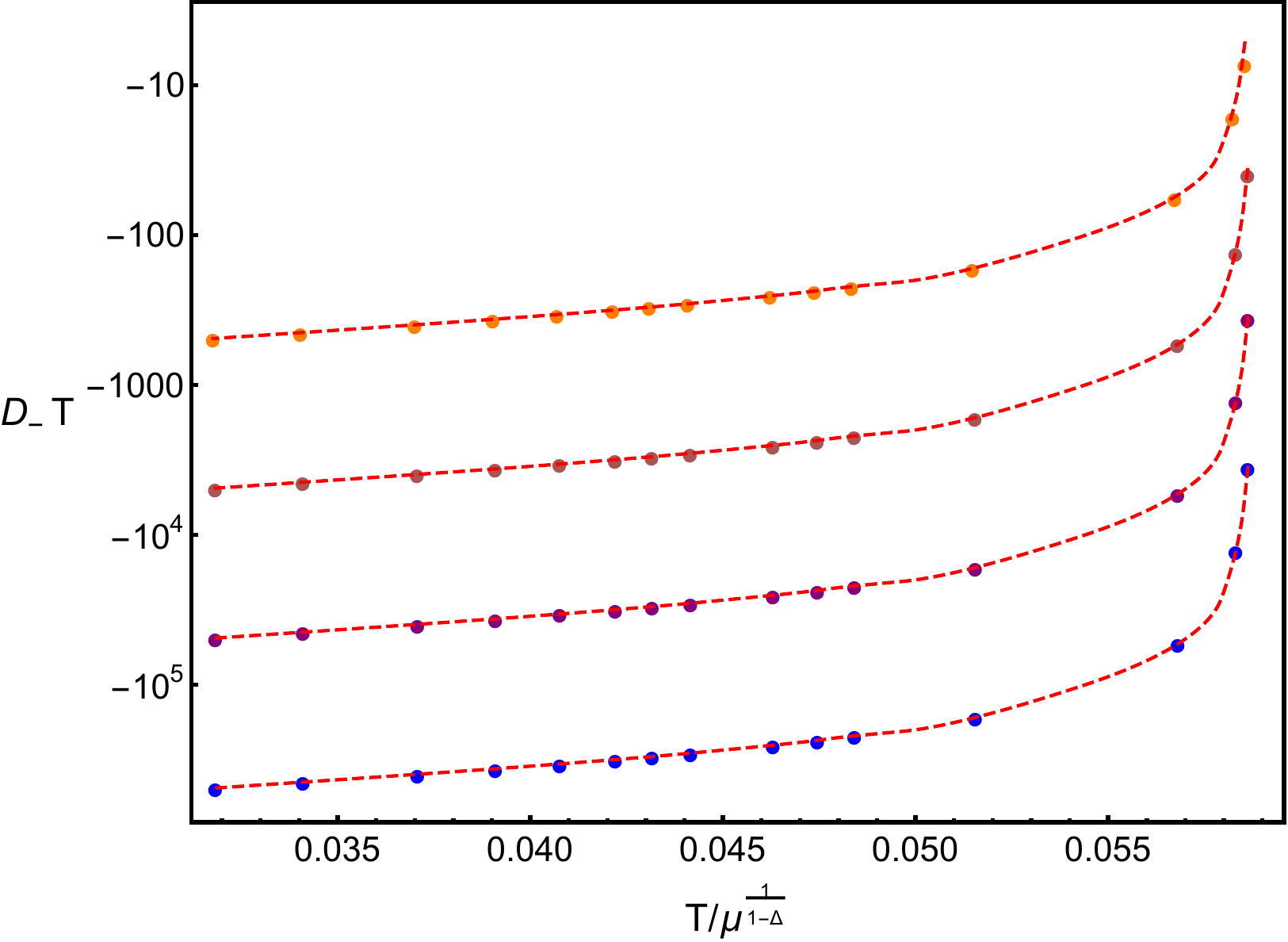}     \caption{Dots mark the dimensionless $k^2$-coefficients entering the dispersion relations of the two lowest quasi-normal modes, extracted for different amounts of explicit breaking. Dashed lines are hydrodynamic predictions \eqref{didi} where transport coefficients and susceptibilities are computed in the purely spontaneous background. Both quantities are computed for the same dimensionless temperature $T/\mu^{1/(1-\Delta_A)}$. Colours indicate different values of the breaking parameter $M^2_A$ -- blue is $10^{-6}$, purple is $10^{-5}$, light brown is $10^{-4}$, and orange is $10^{-3}$. \textbf{Left:} Diffusion constant of gapless mode. \textbf{Right:} The quadratic coefficient of the damped mode.}
    \label{fig7}
\end{figure}

Moving on, we consider the quasi-normal modes at non-zero momentum. Including the leading finite momentum contribution \eqref{fufu} the generalised hydrodynamic dispersion relations for this setup read
\begin{equation}\label{didi0}
    \omega(k)\,=\,-i D_+ k^2 +\dots\,,\qquad \omega(k)=-i \,\Gamma- i\, D_- \,k^2 +\dots,
\end{equation}
where 
\begin{align}\label{didi}
    &D_+=\zeta_3\frac{\rho_s}{\mu}\,+\,\frac{1}{\Gamma}\frac{\rho_s}{\mu \chi_{\rho\rho}}\,,\qquad D_-=\frac{\sigma_0}{\chi_{\rho\rho}}-\frac{1}{\Gamma}\frac{\rho_s}{\mu \chi_{\rho\rho}}\,.
\end{align}
The mode that becomes damped is the charge diffusion of the spontaneously broken phase, while the Goldstone diffusion remains gapless. In figure \ref{fig7} we compare quasi-normal mode data to the $k^2$-coefficients in equation \eqref{didi}; for $D_\pm$ we have used values extracted from the purely spontaneous regime for all quantities except $\Gamma$ -- this is tantamount to considering the parameters $D_\pm$ only at leading order in $M_A$.

In figure \ref{fig7a} we show the dispersion relations of the low-lying quasi-normal modes together with the hydrodynamic expressions in equations \eqref{didi0} and \eqref{didi} in the right-hand panel. For small dimensionless momentum, $k/T \ll 1$, and small values of the breaking (the gauge field mass) the formulae \eqref{didi} are in good agreement with the quasi-normal modes. 

Beyond the small momentum  limit the modes in figure \ref{fig7a} display a feature which is commonly referred to as a $k$-gap \cite{BAGGIOLI20201} and has been observed in several previous works, see for instance \cite{Jimenez-Alba:2014iia,Baggioli:2018nnp,Baggioli:2020loj,Baggioli:2019aqf,Mondkar:2021qsf,Davison:2014lua,Grozdanov:2018ewh}. When increasing the momentum the hydrodynamic diffusive mode and the first damped mode collide at a real value of the momentum $k_*$, sometimes called the critical momentum, and a finite real part develops. These dynamics can be understood directly from the symmetry breaking pattern using the framework of quasi-hydrodynamics \cite{Grozdanov:2018fic}.\footnote{Notice once more the similarities with the dynamics of collective modes in systems with a $U(1)_R \times U(1)_L$ symmetry \cite{Jimenez-Alba:2014iia}, see for example figure 1 in \cite{Stephanov:2014dma}. In that case the original sound mode is the chiral magnetic wave arising due to the axial anomaly and the presence of a background magnetic field.} The dynamics of the $k$-gap has been used to quantify the breakdown of linearised hydrodynamics\footnote{More precisely, the $k$-gap dynamics corresponds to a simple scenario where the critical momentum $k_*$ is real. In general, the breakdown of hydrodynamics must be determined by considering the wave-vector in the complex plane \cite{Grozdanov:2019uhi}.} \cite{Arean:2020eus,Wu:2021mkk,Jeong:2021zsv,Liu:2021qmt,Huh:2021ppg} or its convergence properties \cite{Grozdanov:2019uhi}, as well as to impose a possible universal bound on diffusion \cite{Hartman:2017hhp,Baggioli:2020ljz}. The critical parameters $k_*$ and $\omega_*$ determine at which point linearised hydrodynamics breaks down \cite{Grozdanov:2019kge,Grozdanov:2019uhi}. In our model \eqref{eq:action2} the critical momentum $k_*$ is real and may therefore be extracted directly from the dispersion relations of the two lowest quasi-normal modes.


\section{Discussion}
In this work we have considered the pseudo-spontaneous breaking of a global $U(1)$ symmetry. We constructed a relativistic hydrodynamic theory capable of capturing the effects of charge and phase relaxation arising from pseudo-spontaneous breaking of $U(1)$ symmetry. From the holographic perspective, we built two distinct models where the dual boundary field theories exhibit pseudo-spontaneous $U(1)$ symmetry breaking. We noted that the two models display different types of phase transitions: one which is smeared and one which is sharp. We have shown the validity of our hydrodynamic framework, in the probe limit, by matching the predicted dispersion relations to the low-lying quasi-normal modes of our holographic models, as well as the different frequency and momentum dependent Green's functions. We have observed the emergence of an effective phase relaxation rate due to the interplay between explicit and spontaneous symmetry breaking in holography. In equation \eqref{uni1} we presented a novel, $U(1)$-specific version of the previously proposed universal relation governing the effective phase relaxation as well as provided numerical evidence in its favour, for small amounts of explicit symmetry breaking. Furthermore, we have shown that a non-zero pinning frequency and phase relaxation result in a finite AC and DC conductivity which does not display Drude-like behaviour.\footnote{This is analogous to what was found in the context of translations in the frequency dependent viscosity \cite{Ammon:2019wci}.}

The effective phase relaxation appears in the model considered in section \ref{sec1} but not in the model of section \ref{sec2}, in spite of both models breaking the $U(1)$ symmetry of the boundary theory pseudo-spontaneously. The reason for this may be understood by considering the spontaneously broken $U(1)$ symmetry separately from the shift symmetry inherent to the Goldstone boson. In the model of section \ref{sec1}, the charged scalar source explicitly breaks the boundary $U(1)$ symmetry as well as the Goldstone shift symmetry; this model displays pseudo-Goldstones as well as an effective phase relaxation. The model of section \ref{sec2}, with a massive bulk gauge field, only breaks the $U(1)$ symmetry of the boundary theory but not the global shift of the Goldstone; hence, the Goldstone remains massless and no phase relaxation mechanism emerges. As a direct consequence of this symmetry breaking pattern the spectrum of the second model contains a hydrodynamic diffusive mode. The difference between our models may also be understood by re-expressing the implicit topological symmetry of the Goldstone degrees of freedom as the conservation of an emergent global higher-form symmetry \cite{Grozdanov:2016tdf,Delacretaz:2019brr,Grozdanov:2018ewh,Grozdanov:2018fic,baggioli2021deformations}; in this language, the appearance of a phase relaxation term is related to the explicit breaking of such a symmetry and it appears in the model considered in section \ref{sec1} but not in the model of section \ref{sec2}. Relating the holographic models considered here to those studied in the context of pseudo-spontaneous translational symmetry breaking, the sourced scalar model shows similarities to axionic models \cite{Andrade:2013gsa,Baggioli:2021xuv} or Q-lattices \cite{Donos:2013eha} while the massive-gauge field model is akin to the two-sector models in \cite{Donos:2019txg,Amoretti:2019kuf,Donos:2019hpp} where momentum is relaxed but the Goldstone is not.

We therefore conclude that the effective phase relaxation arises in our model in section \ref{sec1}, but likely even more generally, as a consequence of the explicit breaking of the Goldstone shift symmetry (or equivalently the global higher-form symmetry) and that it is independent of the spontaneously broken global symmetry. This is further confirmed by the appearance of the same emergent phase relaxation mechanism in the context of non-abelian symmetries \cite{Grossi:2020ezz}.

It would be of interest to identify realistic scenarios where the signature of the induced phase relaxation could be detected. A promising avenue could be phason dynamics in incommensurate crystals \cite{Currat2002,Baggioli:2020haa}, where some experiments related to this phenomenon have been done \cite{ollivier1998light,PhysRevLett.107.205502}. Moreover, the case of a non-conserved $U(1)$ axial symmetry due to a topological charge density leads to a plethora of interesting phenomenological features including sphaleron relaxation \cite{Kapusta:2020qdk} and damping of the chiral magnetic wave \cite{Shovkovy:2018tks}. Furthermore, it would be interesting to compute the second order contributions to equilibrium transport, as initiated in~\cite{Grieninger:2021rxd,Kovtun:2018dvd}. 

Potential extensions of our findings regard light dilatons arising from the pseudo-spontaneous breaking of scale invariance \cite{Megias:2014iwa,Elander:2020csd,Elander:2012fk} -- with potential implications for cosmology and beyond standard model physics \cite{Campbell:2011iw,CONTINO2003148} -- and pseudo-magnons or gapped spin waves in magnetic systems \cite{PhysRevLett.121.237201}. One could also check if the universal relation described in this work is valid for type II or type B pseudo-Goldstone bosons which were previously observed in \cite{Amado:2013xya} and \cite{Baggioli:2020edn}. The holographic models of \cite{Amado:2013xya,Baggioli:2020edn,Donos:2021ueh,Amoretti:2021lll} may provide the circumstances for testing these ideas -- explorations in this direction are on-going.


\subsection*{Acknowledgments}
We thank Aristomenis Donos, Blaise Gout\'eraux, Sa\v{s}o Grozdanov, Akash Jain, Karl Landsteiner, Napat Poovuttikul and Oriol Pujolas for useful discussions. We also thank Amadeo Jim\'enez Alba for collaboration at an early stage of this project. S. Gray thanks IFT Madrid for hospitality during the initial stages of this work. M.A. is funded by the \emph{Deutsche Forschungsgemeinschaft} (DFG, German Research  Foundation) under Grant No.\,406235073 within the Heisenberg program. D.A. and S. Grieninger are supported by the `Atracci\'on de Talento' program (2017-T1/TIC-5258, Comunidad de Madrid) and through the grants SEV-2016-0597 and PGC2018-095976-B-C21. M.B. acknowledges the support of the  Shanghai Municipal Science and Technology Major Project (Grant No.2019SHZDZX01). The work of S. Gray is funded by the \emph{Deutsche Forschungsgemeinschaft (DFG)} under Grant No.\,406116891 within the Research Training Group RTG\,2522/1.

\appendix


\section{Hydrodynamics with Energy-momentum Tensor}\label{sec:hydroApp}
In this appendix we compute and present the complete set of hydrodynamic dispersion relations for a system with pseudo-spontaneous breaking of $U(1)$ symmetry -- in two spatial dimensions -- as well as review the superfluid dynamics based on \cite{Herzog:2011ec}. The main result of this appendix is that the `sum relation' \eqref{sumruleprobe} also holds beyond the probe limit. We assume conformal invariance throughout most of this section; however, when explicit breaking is included the conformal nature is only preserved if the deformation used to explicitly break the $U(1)$ symmetry is exactly marginal -- we will consider such a scenario for simplicity.

\subsection{Superfluid Modes}
The complete superfluid hydrodynamics is scoped by supplementing the conservation equation \eqref{conservationProbeSuper} with the conservation of energy and momentum; the hydrodynamic equations are thus
\begin{subequations}\label{conservationSuperApp}
\begin{align}
    \partial_\mu\braket{ T^{\mu\nu}} &= 0, \\
    \partial_\mu\braket{ J^\mu} &= 0,
\end{align}
\end{subequations}
where $\braket{T^{\mu\nu}}$ is the energy-momentum tensor and $\braket{J^\mu}$ is the $U(1)$ current. The dynamics of the Goldstone $\varphi$ is governed by the Josephson relation
\begin{equation}\label{josephsonApp}
    u^\mu \xi_\mu = -\mu,
\end{equation}
where $\xi_\mu = \partial_\mu \varphi$ for the same reasons as in the main text, $u^\mu$ is the normal fluid velocity normalised such that $u^\mu u_\mu = -1$ and $\mu$ is the chemical potential.

Up to first order in derivatives the constitutive relations and the spatial derivative of the Josephson relation read 
\begin{subequations}\label{eq:constrelationSuperApp}
\begin{align}
\begin{split}
   \braket{T^{\mu\nu}} &= \varepsilon \,u^\mu u^\nu + p\,\Delta^{\mu\nu} + 2\rho_s\,\mu \,n^{(\mu}u^{\nu)} + \rho_s\, \mu\, n^{\mu}n^\nu -\eta \,\sigma^{\mu\nu} + \mathcal{O}(\partial^2) , 
\end{split}
    \\
    \braket{J^{\mu}} &= \rho_t~ u^\mu + \rho_s~ n^\mu - \sigma_0 ~T~ \Delta^{\mu\nu} \partial_\nu \del{\frac{\mu}{T}} + \mathcal{O}(\partial^2),\\
                u^\mu \partial_\mu \del{\tensor{\Delta}{_\nu^\rho}\xi_\rho} &= \tensor{\Delta}{_\nu^\rho}\left[-\partial_\rho\mu + \zeta_3\, \partial_\rho \partial_\mu \del{\rho_s n^\mu} \right]+ \mathcal{O}(\partial^2) , \label{josephsonDerivApp}
\end{align}
\end{subequations}
where $\Delta^{\mu\nu}=u^\mu u^\nu + \eta^{\mu\nu}$ is a projector with $\eta^{\mu\nu}$ being the Minkowski metric; $\varepsilon$ is the energy density; $T$ denotes the temperature; $p$ is the thermodynamic pressure; the normal and superfluid charge densities are denoted $\rho_n$ and $\rho_s$ respectively; the total charge density is given by $\rho_t =\rho_n + \rho_s$; and $n^\mu$ is the superfluid velocity defined by \eqref{supervelocity}.
The transport coefficients entering in \eqref{eq:constrelationSuper} are the shear viscosity $\eta$, conductivity $\sigma_0$, and superfluid bulk viscosity $\zeta_3$; $\sigma_0$ and $\zeta_3$ are given by the Kubo formulae \eqref{kuboSuper}, and $\eta$ is given by
\begin{align}
    \eta &= -  \lim_{\omega\to0}\frac1\omega \mathrm{Im}\left[G^\mathrm{R}_{T^{xy}T^{xy}}(\omega,0) \right], 
\end{align}
where $G^\mathrm{R}_{ab}(\omega,\mathbf{k})$ is the retarded Green's function, $\omega$ denotes the frequency and $\mathbf{k}$ is the spatial momentum vector which will taken to be $\mathbf{k} = (k,0)$. 

The longitudinal hydrodynamic modes can be calculated from the conservation equations \eqref{conservationSuperApp} together with the constitutive relations and Josephson relation \eqref{eq:constrelationSuperApp}. We find four hydrodynamic modes in a configuration of two pairs of propagating sound modes with dispersion relations
\begin{equation}\label{hydrodispApp}
    \omega(k) = \pm v_n k - \frac{i}{2} \Gamma_n k^2 + \dots,
\end{equation}
where the ellipsis denotes terms at higher order in $k$. These two modes are historically defined as first and second sound \cite{doi:10.1063/1.3248499}. The speed and attenuation of first sound are given by
\begin{equation}
    v_1^2 = \frac{1}{2}, \qquad \Gamma_1 = \frac{\eta}{w},
\end{equation}
where $s$ is the entropy and $w = sT + \mu\rho_t$ the total enthalpy density, also referred to as the momentum susceptibility. Note that the speed of first sound is fixed by conformal invariance. Similarly, the same parameters for second sound are 
\begin{align}
    v_2^2 = \frac{\rho_s \sigma^2}{(\partial \sigma/\partial T)_\mu w}, \qquad \Gamma_2 = \sigma_0 \frac{\mu w}{(\partial \sigma/\partial T)_\mu T^2 \rho_t^2} + \eta \frac{\mu \rho_s}{w h} + \zeta_3 \frac{\rho_s w}{\mu h},
\end{align}
where $h = sT + \mu \rho_n$ is the enthalpy density of the normal component. In accordance with \cite{Herzog:2011ec}, we have for convenience introduced the reduced entropy $\sigma=s/\rho_t$ and used the conformal scaling form of the pressure $p=T^3 f(\mu/T)$ to express thermodynamic derivatives in terms of $\sigma$ via
\begin{subequations}\label{thermosubs}
\begin{align}
    \del{\frac{\partial \rho_t}{\partial T}}_\mu &= \frac{\partial^2 p}{\partial T\partial\mu} = \del{\frac{\partial s}{\partial \mu}}_T = \frac{\rho_t}{w}\left[2s - \rho_t T \del{\frac{\partial \sigma}{\partial T}}_\mu \right], \\
    \del{\frac{\partial \rho_t}{\partial \mu}}_T &=  \frac{\rho_t}{w }\left[2s + \rho_t T^2 \del{\frac{\partial \sigma}{\partial T}}_\mu \right],\\
   \del{ \frac{\partial s}{\partial T}}_\mu &= \frac{2(sT-\mu\rho_t)}{T^2} + \frac{\mu^2}{T^2} \frac{\rho_t}{w}\left[2\rho_t + \frac{\rho_t T^2}{\mu} \del{\frac{\partial \sigma}{\partial T}}_\mu \right],
\end{align}
\end{subequations}
where the subscript on the derivatives denotes the quantity held fixed.

\subsection{Broken Superfluid Modes}
When the $U(1)$ symmetry is also explicitly broken the current conservation becomes modified as in equation \eqref{chargeNonCons}; the hydrodynamic equations are thus
\begin{subequations}\label{hydroeqapp}
\begin{align}
    \partial_\mu\braket{ T^{\mu\nu}} &= 0,\\
    \partial_\mu\braket{J^\mu} &= \Gamma u_\mu \braket{J^\mu} + m\,\varphi,
\end{align}
where $\Gamma$ denotes the charge relaxation rate and $m$ is related to the explicit breaking parameter. The relaxed Josephson relation reads 
\begin{equation}
    (u^\mu\partial_\mu + \Omega)\tensor{\Delta}{_\nu^\rho}\xi_\rho =\tensor{\Delta}{_\nu^\rho}\left[ -\partial_\rho\mu + \zeta_3 \partial_\rho \partial_\mu \del{\rho_s n^\mu}\right] + \mathcal{O}(\partial^2),  \label{relaxedJoeapp} 
\end{equation}
\end{subequations}
where $\Omega$ is the phase relaxation. 

Using the constitutive relations \eqref{eq:constrelationSuperApp} and hydrodynamic equations \eqref{hydroeqapp} we may calculate the hydrodynamic modes using standard methods.\footnote{Note that we in this setup (as opposed to the probe limit in the main text) consider fluctuations of all hydrodynamic variables.} The analysis yields a pair of propagating modes with dispersion relation
\begin{equation}\label{prophydrobroken}
    \omega(k) = \pm v_1 k - \frac{i}{2}\Gamma_1 k^2 + \dots,
\end{equation}
with speed and attenuation of sound
\begin{equation}
    v_1^2 = \frac{1}{2}, \qquad \Gamma_1 = \frac{\eta}{h}.
\end{equation}
The above mode behaves exactly as the propagating longitudinal mode in a normal conformal fluid. If we temporarily revoke the conformal invariance the speed of sound of the mode \eqref{prophydrobroken} is given by
\begin{equation}
    v_{nc}^2 = \frac{m s +(s \tilde{\chi}_{\rho_t \mu} - \rho_t \tilde{\chi}_{\rho_t T}) \Gamma \Omega }{m (T \tilde{\chi}_{s T} + \mu \tilde{\chi}_{\rho_t T}) + T(\tilde{\chi}_{sT}\tilde{\chi}_{\rho_t \mu} - \tilde{\chi}_{s\mu} \tilde{\chi}_{\rho_t T})\Gamma\Omega } ,
\end{equation}
where $\tilde{\chi}_{ab}$ denotes the thermodynamic derivative of $a$ with respect to $b$, as in the left-hand sides of equations \eqref{thermosubs}. The non-conformal sound attenuation can be also derived but its expressions is rather lengthy and therefore not presented.

Moreover, the spectrum contains a pair of gapped and damped modes with dispersion relation
\begin{equation}\label{gappedDiffApp}
    \omega(k) = \tilde{\alpha}_{\pm} - i \tilde{D}_{\pm} k^2 + \dots, 
\end{equation}
where 
\begin{equation}\label{gapapp}
   \tilde{\alpha}_\pm = -\frac{i}{2}(\Gamma + \Omega) \pm \frac{1}{2}\sqrt{4 \,\tilde{\omega}_0^2-(\Gamma - \Omega)^2 }, \qquad \tilde \omega_0^2\equiv  \frac{m\,s\,\mu}{(\partial\sigma/\partial T)_\mu T \rho_t^2}.
\end{equation}
The complete expression for the coefficient $\tilde{D}_\pm$ is rather cumbersome to be presented in full generality, however for $\Gamma = \Omega = 0$ it reads 
\begin{equation}\label{Dapp}
    \tilde{D}_\pm = \sigma_0  \frac{\mu w}{2(\partial\sigma/\partial T)_\mu T^2\rho_t^2}+\zeta_3\frac{ \rho_s w}{2\mu h}\pm \frac{i}{2}\frac{s^{3/2} \sqrt{T}\rho_s }{h \rho_t \sqrt{(\partial\sigma/\partial T)_\mu m\mu} }.
\end{equation}

We may at this point make an observation with regards to the coefficients of the $k^2$-terms in the dispersion relations presented in this appendix; namely, they obey the sum relation
\begin{equation}\label{sumruleApp}
   \tilde{D}_+ + \tilde{D}_- + \Gamma_0 = \Gamma_1 + \Gamma_2,
\end{equation}
where the left-hand side contains the coefficients in the pseudo-spontaneous regime while the right-hand side is purely spontaneous. The relation \eqref{sumruleApp} also holds for non-zero $\Gamma$ and $\Omega$. See equation \eqref{sumruleprobe} in the main text for the probe-limit equivalent to equation \eqref{sumruleApp}.\footnote{An analogous sum relation also holds in the case of pseudo-spontaneous breaking of translational invariance.}

%
%
\section{Note on the Universal Relation} 
\label{uniapp}
In this appendix we will provide further evidence for the universal appearance of the effective phase relaxation. To this end, we consider the  dispersion relation in equation (4.11) of \cite{Donos:2021pkk} at zero momentum\footnote{This expression is derived in \cite{Donos:2021pkk} by using perturbative analytic methods in the holographic model of section \ref{sec1} at zero chemical potential $\mu=0$.} -- in their notations, this reads 
\begin{equation}\label{gapforUni}
    \omega\,=\,\sqrt{\frac{w}{\chi_{QQ}}}-\frac{i}{2}\,\frac{w\,\tilde \Xi}{\chi_{QQ}^2},
\end{equation}
where $w$ denotes the infinitesimal explicit breaking parameter. Comparing the above expression to our hydrodynamic result \eqref{pred1}, we may identify 
\begin{equation}\label{toge}
    \omega_0^2 \equiv \frac{w}{\chi_{QQ}}\,,\qquad \Omega \equiv \frac{w\,\tilde \Xi}{\chi_{QQ}^2},
\end{equation}
in which terms of order $w^2$ have been consistently neglected in \eqref{gapforUni} in the limit of small explicit breaking. From equations \eqref{toge} we immediately obtain 
\begin{equation}
    \Omega\,=\,\omega_0^2\,\frac{\tilde \Xi}{\chi_{QQ}} \label{don},
\end{equation}
which coincides with the expression \eqref{uni1} in the main text when notations are matched, as well as the original proposal of \cite{Amoretti:2018tzw} in the context of translational symmetry breaking. This brief analysis further supports the universal nature of the effective phase relaxation.

\section{Numerical Methods}\label{app:num}

All backgrounds and solutions to the linearised fluctuation equations are obtained numerically by means of pseudo-spectral methods as we have established in previous work~\cite{Grieninger:2020wsb,Grieninger:2021rxd,Ghosh:2021naw,Arean:2021tks,Ammon:2016fru} (see~\cite{Boyd00} for an introductory textbook). To employ the pseudo-spectral methods, we discretise the radial dependence of the unknown functions on a Chebychev-Lobatto grid with $N$ gridpoints
\begin{equation}
    u_m=\frac12\left(\cos\left(\frac{m\,\pi}{N-1}\right)+1\right), \quad m\in [0,N-1],\label{eq:chebylobatto}
\end{equation} 
and replace all radial derivatives of the unknown functions (in a Chebychev basis) by the corresponding derivative matrices. To obtain the background solutions, we solve the discretised non-linear ordinary differential equation with help of a Newton-Raphson scheme.

The quasi-normal modes are computed by recasting the linearised fluctuation equations about the numerically constructed background solution in terms of a generalised eigenvalue problem with respect to the quasi-normal mode frequency $\omega$,
 \begin{equation}
(\bm{A}\,\omega-\bm{B})\,\bm{x}_1=0,
 \end{equation}
 where $\bm{A}(k)$ and $\bm{B}(k)$ are differential operators consisting of the discretised fluctuation equations and $\bm{x}_1$ is given by the fluctuation vector consisting of the gauge- and scalar field fluctuations $\{a_t,\,a_x,\delta,\,\eta\}$ on the gridpoints of the Chebychev-Lobatto grid.
 
 \begin{figure}
    \centering
    \includegraphics[width=0.45\linewidth]{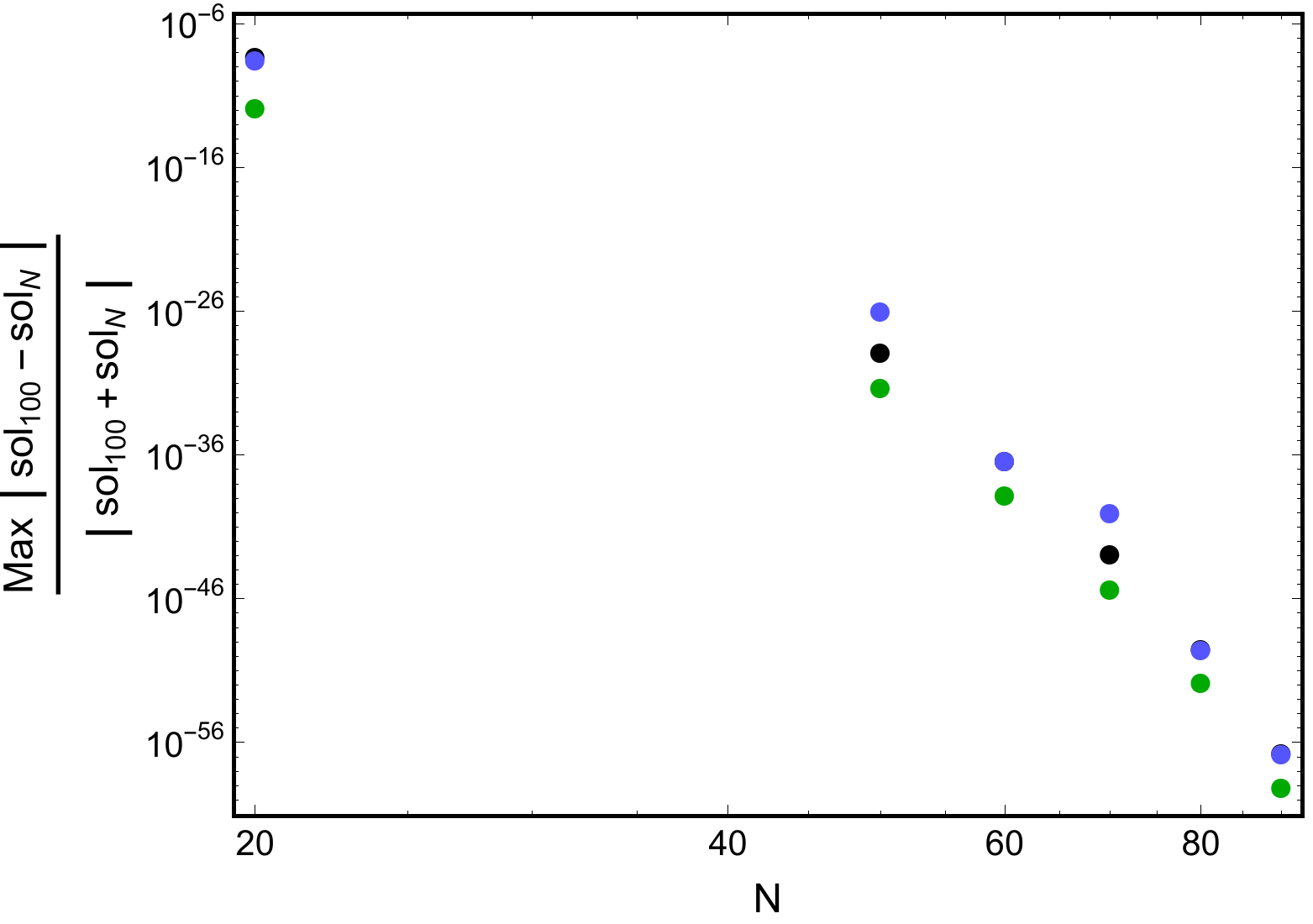}\qquad \includegraphics[width=0.45\linewidth]{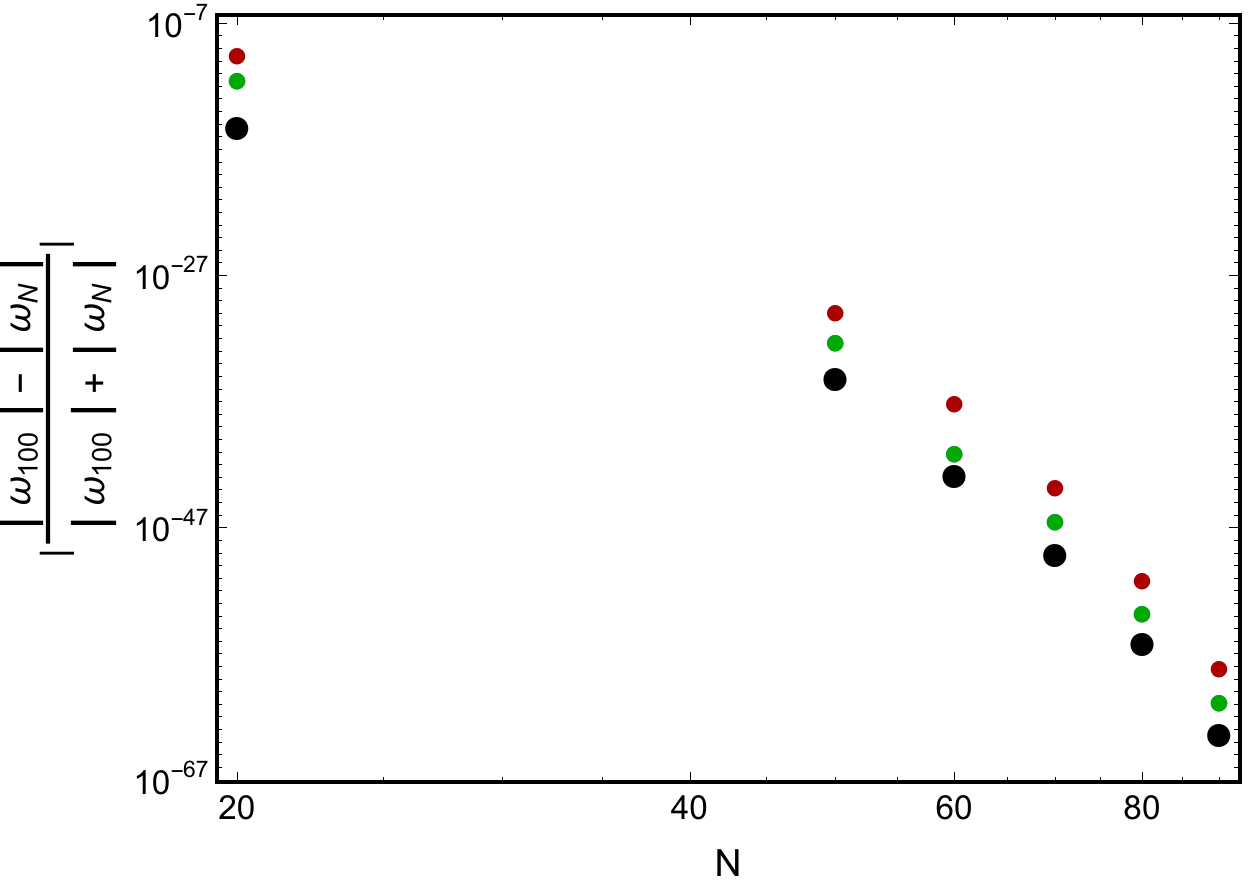}     \caption{\textbf{Left:} Relative supremums norm of the numerical solution for the background functions $\psi_1$ (black),\,$\psi_2$ (blue),\, $A_t$ (green) at $\langle \mathcal O\rangle/T^2=5.6147$ and $\alpha/T=10^{-4}$. \textbf{Right:} Relative error of the lowest QNMs for the background of the left plot and $k/T=0.4189$.}
    \label{conv1}
\end{figure}

In the following, we will briefly discuss the convergence of our numerical scheme for the holographic models used in section~\ref{sec1} and~\ref{sec:four}.

\begin{figure}
    \centering
    \includegraphics[width=0.45\linewidth]{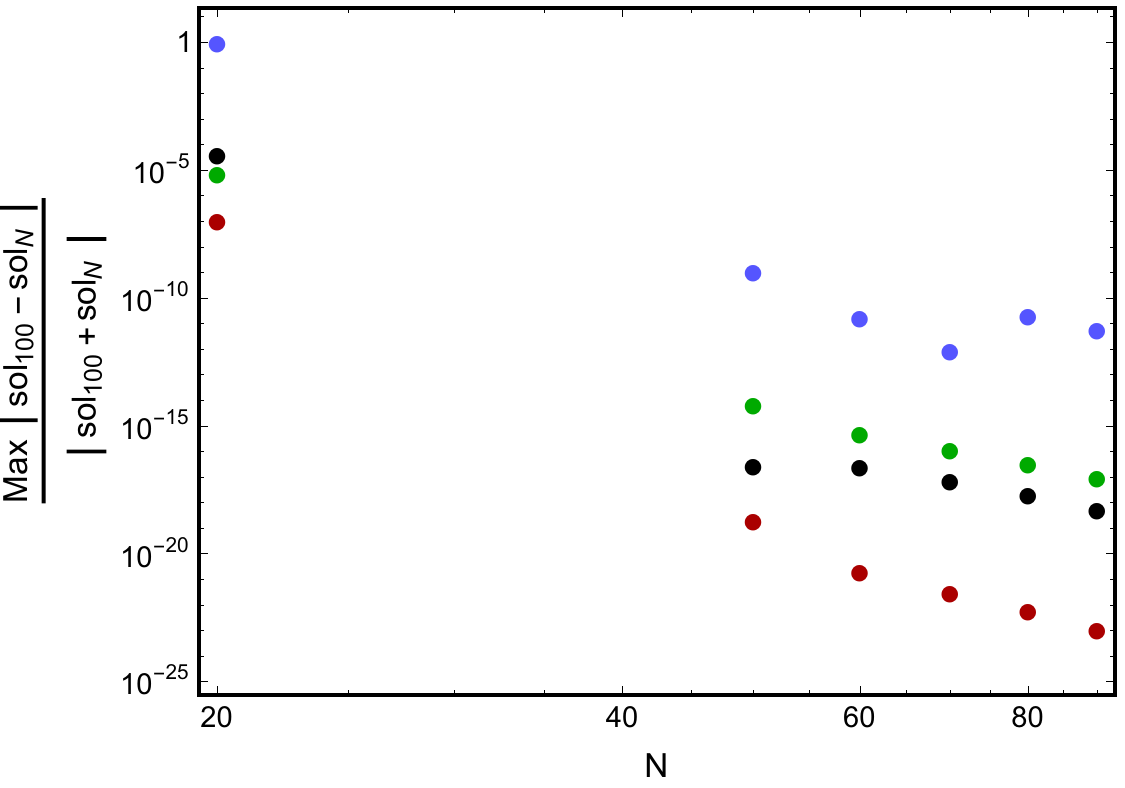}\qquad \includegraphics[width=0.45\linewidth]{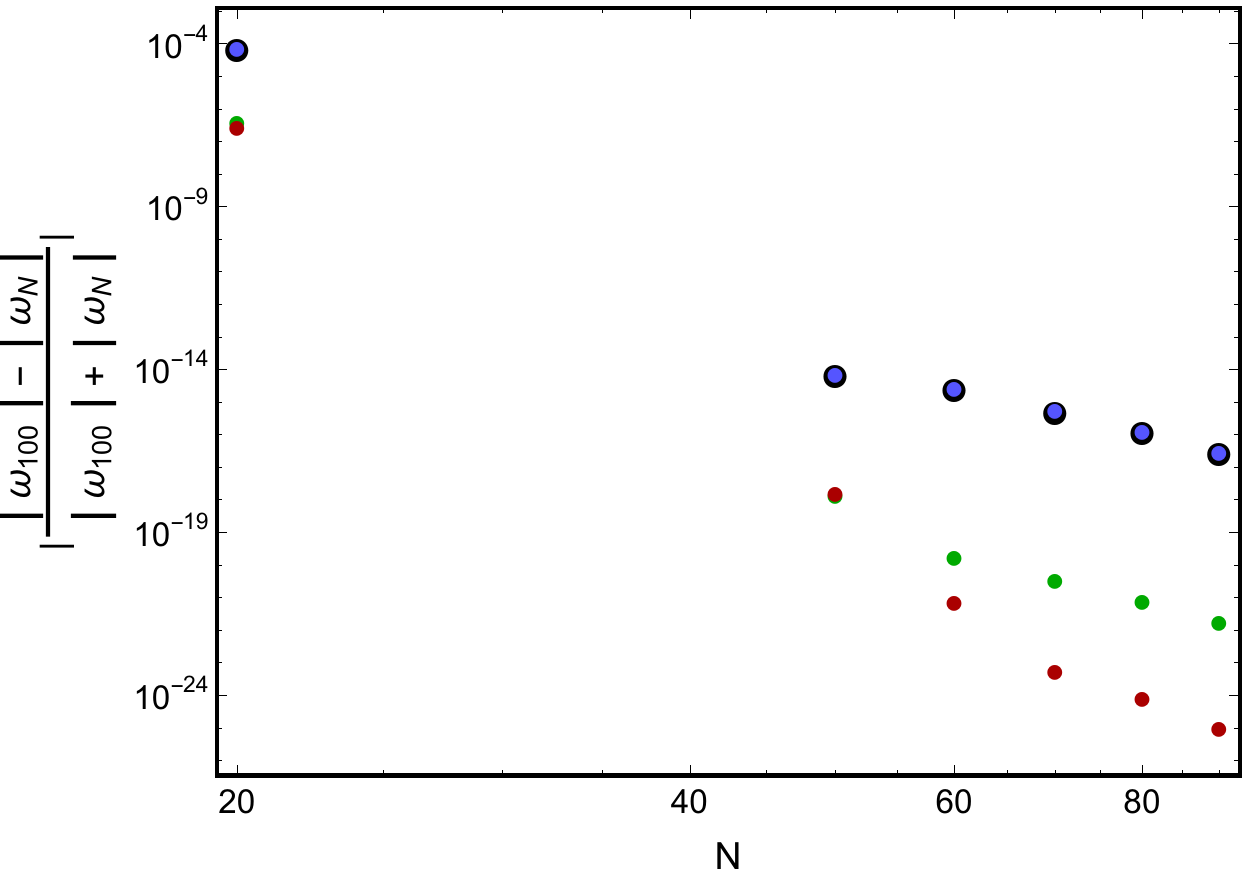}     \caption{\textbf{Left:} Relative supremums norm of the numerical solution for the background functions $\psi_1$ (black),\,$\psi_2$ (blue),\, $A_t$ (green),\, $\theta$ (red) at $\langle \mathcal O\rangle/T^2=7.301$ and $M^2=10^{-6}$. \textbf{Right:} Relative error of the lowest QNMs for the background of the left plot and $k/T=4.189\cdot10^{-6}$.}
    \label{conv2}
\end{figure}

For the background solutions, we provide proof that the numerical solution converges when we increase the number of gridpoints. In the left plot of figure \ref{conv1}, we show the supremums norm of a solution with a small number of gridpoints compared to a high resolution solution ($N=100$). We compare both solutions on an equidistant grid (not the Chebychev-Lobatto grid!) with 101 gridpoints. As is evident from the plot, the supremums norm decreases for an increasing number of spectral gridpoints. In the right plot we provide evidence that the relative error of the four lowest quasi-normal modes, as compared to a high resolution solution, converges to zero. In figure \ref{conv2}, we provide an analogous plot for the model of section~\ref{sec:four}.
 
\bibliographystyle{JHEP}
\bibliography{refs.bib}
\end{document}